\documentstyle[graphicx,preprint,aps]{revtex}

\begin{document}
\draft
%
%
\title{Anomalous Gauge Boson Couplings in the 
         $\bbox{\mathrm{e^+ e^-} \rightarrow \mathrm{Z}\mathrm{Z}}$ Process}
\author{J. Alcaraz and M.A. Falag\'an 
        \thanks{partially supported by CICYT Grant: AEN96-1645}}
\address{CIEMAT, Avda. Complutense 22, 28040-MADRID, Spain}
\author{E. S\'anchez}
\address{CERN, 1211 Gen\`eve 23, Switzerland} 

\date{\today}
\maketitle

\begin{abstract}
We discuss experimental aspects related to the 
$\mathrm{e^+ e^-} \rightarrow \mathrm{Z}\mathrm{Z}$ process
and to the search for anomalous ZZV couplings 
(V$= \mathrm{Z}, \gamma$) at LEP2 and future $\mathrm{e^+ e^-}$ colliders. 
We present two possible approaches for a realistic study of the 
reaction and discuss the differences between them. We find that the optimal 
method to study double Z resonant production and to quantify the 
presence of anomalous couplings requires the use of a complete 
four-fermion final-state calculation. 
\end{abstract}


\section*{Introduction}

  Pair production of $\mathrm{Z}$ bosons is one of the new physics processes 
to be studied at LEP2 and at future high energy $\mathrm{e^+ e^-}$ colliders. 
Although it is a process with a rather low cross section (below 1 pb) and 
experimentally difficult to observe (large and almost irreducible backgrounds),
 LEP2 gives the first opportunity to perform a measurement and to look for 
deviations from the Standard Model (SM). In addition, a good understanding of 
the process is necessary, since it is one of the relevant backgrounds in the 
search for the Higgs particle. At future $\mathrm{e^+ e^-}$ colliders, 
with luminosities of the order of 100 \mbox{fb$^{-1}$}, several thousands of 
events will provide stringent tests of the SM.

  The study of triple-gauge boson couplings is one of the key issues at 
present and future colliders. Anomalous 
$\mathrm{Z}\gamma\mathrm{V}$ couplings have been searched for 
at the Tevatron and at LEP \cite{zg}. The first experimental limits on 
anomalous $\mathrm{Z}\mathrm{Z}\mathrm{Z}$ couplings have been provided by 
the L3 Collaboration \cite{zzpapers}.

  The report is organized as follows. First, the SM amplitude is presented 
and the effects of possible anomalous ZZV couplings at LEP2 and at future 
$\mathrm{e^+ e^-}$ colliders are discussed. Second, we describe two
reweighting approaches developed for the search for anomalous couplings at 
LEP2. These approaches are compared and their differences are pointed out.
Finally, the fitting techniques used to quantify the possible presence of 
anomalous couplings are briefly presented.

\section*{Standard Model amplitude for the 
    $\bbox{\mathrm{\lowercase{e}^+ \lowercase{e}^-} \rightarrow 
    \mathrm{Z}\mathrm{Z}}$ process}

\begin{figure}[htbp]
\begin{center}
    \includegraphics[width=8cm]{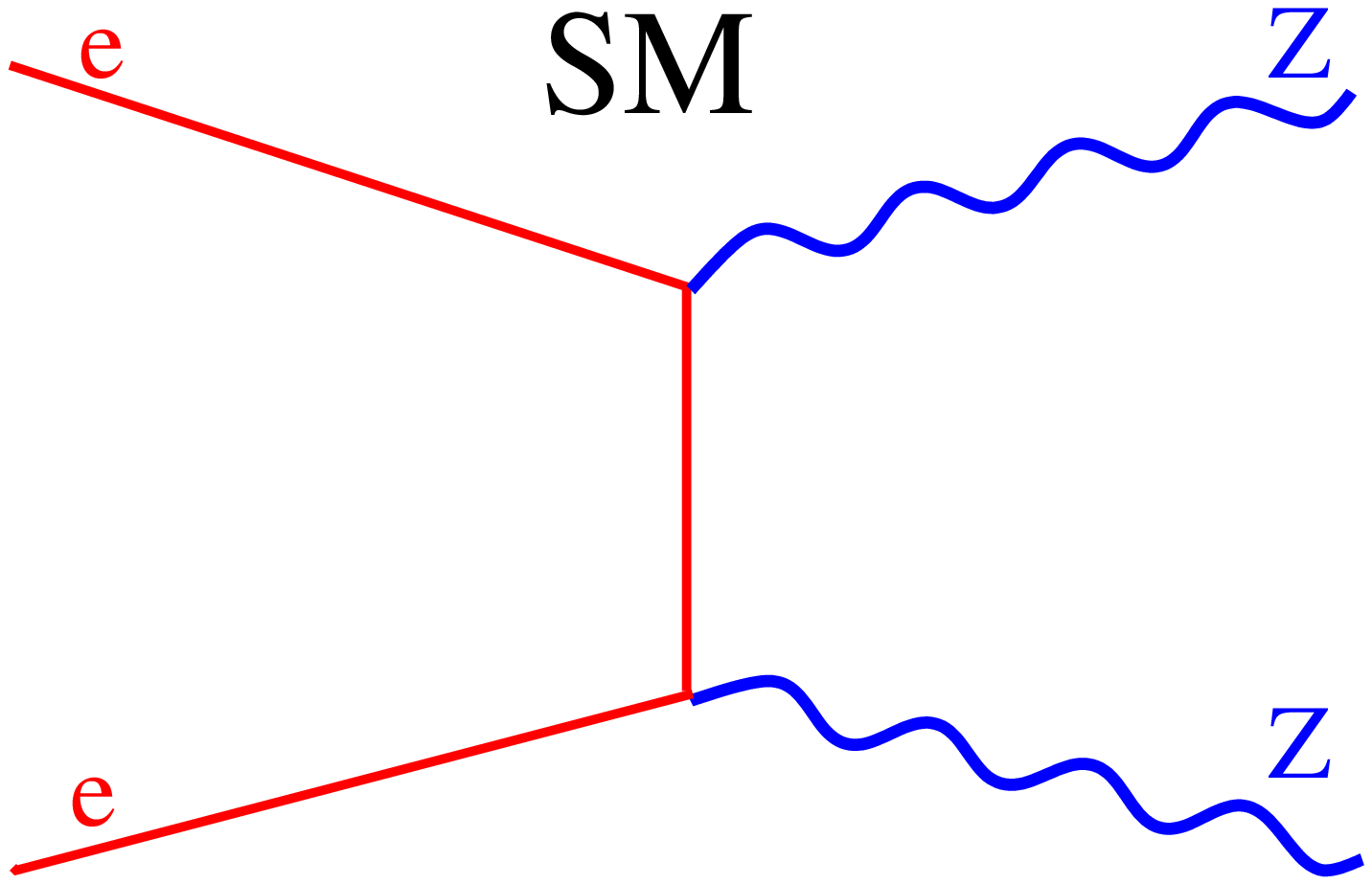}
\end{center}
\begin{center}
    \includegraphics[width=8cm]{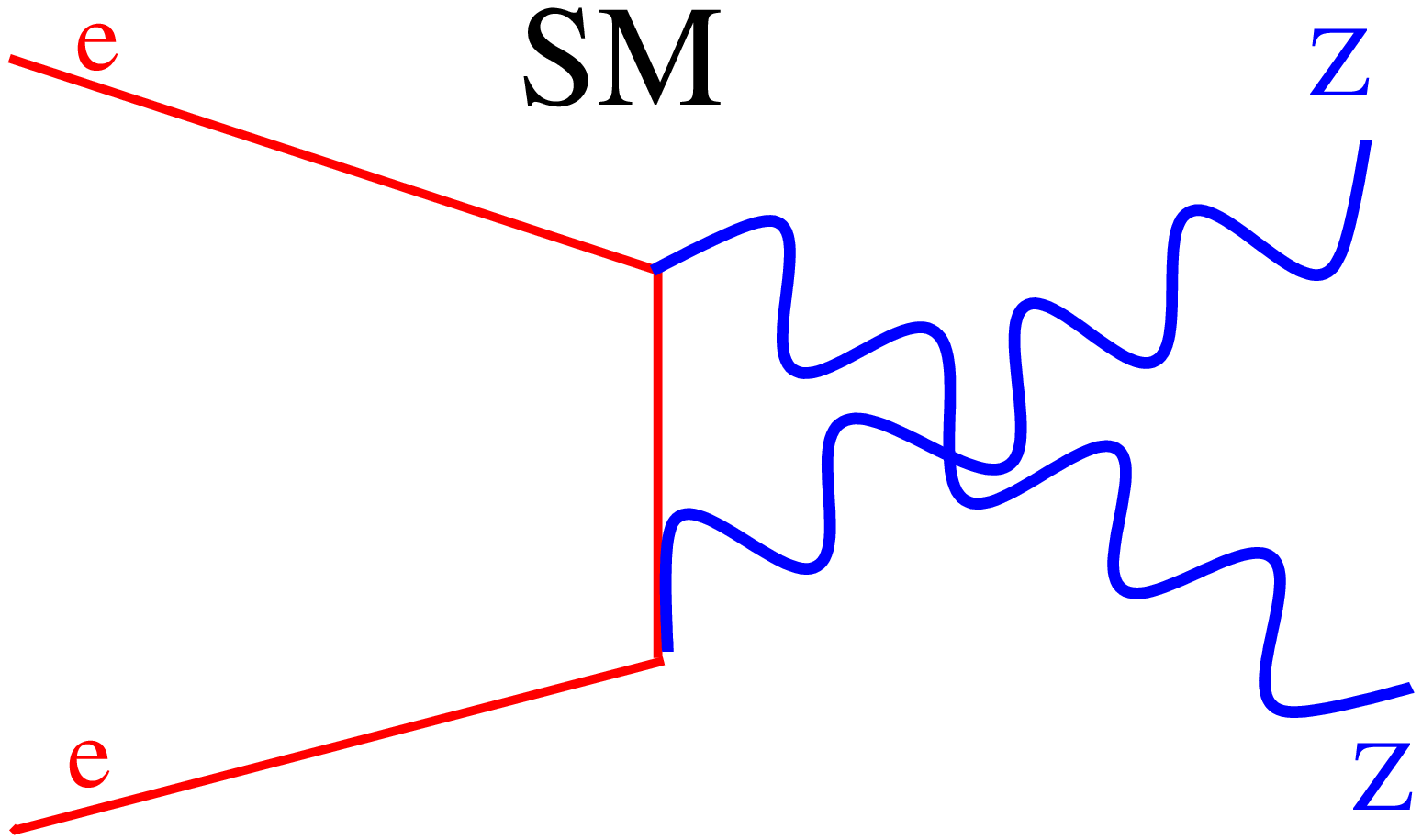}
\end{center}
\caption{ Diagrams contributing at first order to the 
         $\mathrm{e^+ e^-} \rightarrow \mathrm{Z} \mathrm{Z}$ process
         in the Standard Model.}
\label{fig:zzsm}
\begin{center}
    \includegraphics[width=10cm]{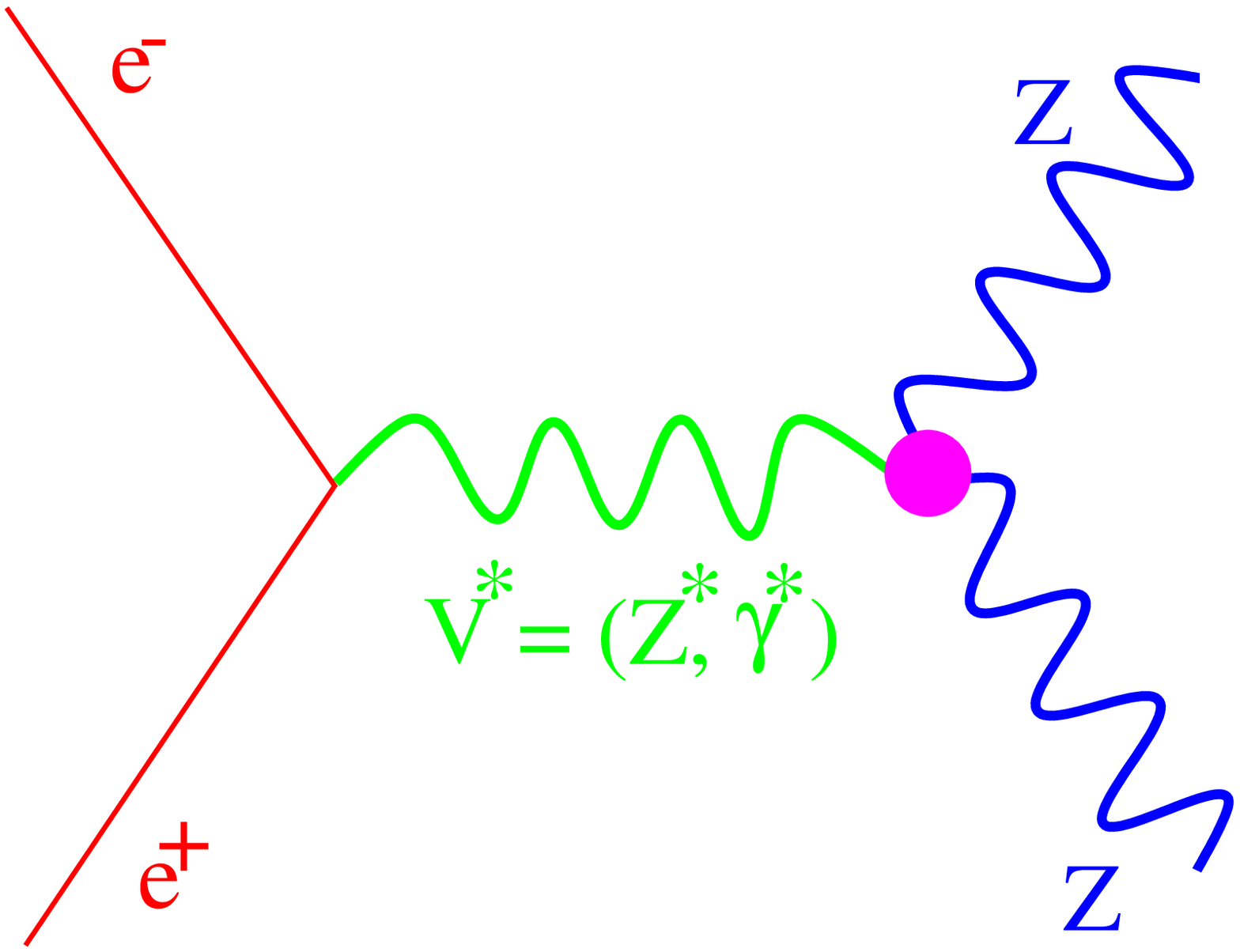}
\end{center}
\caption{ Diagram with anomalous $\mathrm{Z}\mathrm{Z}\gamma$ and 
             $\mathrm{Z}\mathrm{Z}\mathrm{Z}$ couplings contributing to the 
             $\mathrm{e^+ e^-} \rightarrow \mathrm{Z} \mathrm{Z}$ process.}
\label{fig:zzanom}
\end{figure}

   The diagrams contributing at first order to the 
$\mathrm{e^+ e^-} \rightarrow \mathrm{Z} \mathrm{Z}$ process 
in the Standard Model are shown in Figure \ref{fig:zzsm}.  
We will assume a collision
in the center-of-mass system with total energy $\sqrt{s}$ and neglect the 
effect of the electron mass. The following notation is used:

\begin{itemize}
\item Electron four-momentum $k$ and helicity $\sigma$:
   \begin{itemize}
      \item[] $k = (\frac{\sqrt{s}}{2}, 
                  \frac{\sqrt{s}}{2}~\hat{e}_z)$
      \item[] $\sigma \in \{-1,1\}$
   \end{itemize}

\item Positron four-momentum $\overline{k}$ and helicity $\overline{\sigma}$:
   \begin{itemize}
      \item[] $\overline{k} = (\frac{\sqrt{s}}{2},
                -\frac{\sqrt{s}}{2}~\hat{e}_z)$
      \item[] $\overline{\sigma} \in \{-1,1\}$
   \end{itemize}

\item $\mathrm{Z}$ four-momentum $q_{\mathrm {Z_1}}$ and polarization 
        $\epsilon_{\mathrm {Z_1}}$:
   \begin{itemize}
      \item[] $q_{\mathrm {Z_1}} = 
                (E,\sqrt{E^2- M_{\mathrm {Z_1}}^2}~\hat{q})$; \hspace{2cm}
              $E = \frac{\sqrt{s}}{2} + \frac{ M_{\mathrm {Z_1}}^2- 
                      M_{\mathrm {Z_2}}^2}{2\sqrt{s}}$
      \item[] $\epsilon_{\mathrm {Z_1}} \equiv 
            \epsilon_{\mathrm {Z_1}}(\lambda_{\mathrm {Z_1}})$; \hspace{2cm}
            $\lambda_{\mathrm {Z_1}} \in \{-1,0,1\}$
   \end{itemize}

\item $\mathrm{Z}$ four-momentum $q_{\mathrm {Z_2}}$ and polarization 
              $\epsilon_{\mathrm {Z_2}}$:
   \begin{itemize}
      \item[] $q_{\mathrm {Z_2}} = 
  (E^\prime,-\sqrt{E^{\prime~2}- M_{\mathrm {Z_2}}^2}~\hat{q})$; \hspace{2cm}
              $E^\prime = \frac{\sqrt{s}}{2} 
     + \frac{ M_{\mathrm {Z_2}}^2- M_{\mathrm {Z_1}}^2}{2\sqrt{s}}$
      \item[] $\epsilon_{\mathrm {Z_2}} \equiv 
    \epsilon_{\mathrm {Z_2}}(\lambda_{\mathrm {Z_2}})$; \hspace{2cm}
             $\lambda_{\mathrm {Z_2}} \in \{-1,0,1\}$
   \end{itemize}

\end{itemize}

\noindent
where the electron is assumed to collide along the +z axis ($\hat{e}_z$), and 
the $\mathrm{Z}$ -with mass $ M_{\mathrm {Z_1}}$- goes along the direction 
given by $\hat{q} = (\sin\theta_{\mathrm{Z}} \cos\phi_{\mathrm{Z}}, 
\sin\theta_{\mathrm{Z}} \sin\phi_{\mathrm{Z}}, \cos\theta_{\mathrm{Z}})$. 
The masses $ M_{\mathrm {Z_1}}$ and $ M_{\mathrm {Z_2}}$ are not assumed to
be equal to the on-shell mass $ m_{\mathrm{Z}}$ because in the following 
they will be consider as virtual particles decaying into fermions.

The matrix element for the 
$\mathrm{e^+ e^-} \rightarrow \mathrm{Z} \mathrm{Z}$ reaction is determined 
by the same method followed in \cite{hagiwara_et_al}. It reads:
\begin{eqnarray}
M_{\mathrm{Z}\mathrm{Z}}
 (\sigma,\overline{\sigma},\lambda_{\mathrm {Z_1}},\lambda_{\mathrm {Z_2}}) = 
     - (g_\sigma^{\mathrm{Z}\mathrm{e^+ e^-}})^2 \sqrt{s}~
            \delta_{\sigma,-\overline{\sigma}}~
    \left [ 
           \frac{S(\epsilon_{\mathrm {Z_2}}^*,q_{\mathrm {Z_1}}
                   ,\epsilon_{\mathrm {Z_1}}^*,\sigma)}
                {-2 (k q_{\mathrm {Z_1}}) + M_{\mathrm {Z_1}}^2}
           + 
           \frac{S(\epsilon_{\mathrm {Z_1}}^*
                   ,q_{\mathrm {Z_2}}
                   ,\epsilon_{\mathrm {Z_2}}^*,\sigma)}
                {-2 (k q_{\mathrm {Z_2}}) + M_{\mathrm {Z_2}}^2}
    \right ]
\end{eqnarray}

   The functions $S(\epsilon_a,q_b,\epsilon_b,\sigma)$ are given by:
\begin{eqnarray}
S(\epsilon_a,q_b,\epsilon_b,+) & = &
\left ( \begin{array}{lr}
 \epsilon_a^1+i \epsilon_a^2 , & -\epsilon_a^0-\epsilon_a^3 \\ 
        \end{array} \right )
\left ( \begin{array}{cc}
     \sqrt{s}-q_b^0-q_b^3, & - q_b^1 + i q_b^2 \\
     - q_b^1 - i q_b^2, & - q_b^0 + q_b^3 \\ \end{array} \right )
\left ( \begin{array}{c}
    \epsilon_b^0- \epsilon_b^3 \\
    - \epsilon_b^1-i \epsilon_b^2 \\ \end{array} \right ) \\
S(\epsilon_a,q_b,\epsilon_b,-) & = &
\left ( \begin{array}{lr}
 \epsilon_a^0+ \epsilon_a^3 , & \epsilon_a^1-i \epsilon_a^2 \\ 
   \end{array} \right )
\left ( \begin{array}{cc}
     -q_b^0+q_b^3, & q_b^1 - i q_b^2 \\
     q_b^1 + i q_b^2, & \sqrt{s} - q_b^0 - q_b^3 \\ \end{array} \right )
\left ( \begin{array}{c}
    \epsilon_b^1- i \epsilon_b^2 \\
    \epsilon_b^0 - \epsilon_b^3 \\ \end{array} \right )
\end{eqnarray}

\noindent
where the components of the four-vectors are denoted by superscripts.
  The left/right effective couplings of fermions to neutral gauge bosons are 
given by:
\begin{eqnarray}
  g_+^{\mathrm{Z}\mathrm{f}\bar{\mathrm{f}}} & = & -2 Q_{\mathrm{f}} 
\sin^2\!\overline{\theta}_{\mathrm{W}} \left( \sqrt{2} G_\mu  m_{\mathrm{Z}}^2 
                                              \right )^{1/2} \\
  g_-^{\mathrm{Z}\mathrm{f}\bar{\mathrm{f}}} & = & 
     g_+^{\mathrm{Z}\mathrm{f}\bar{\mathrm{f}}} + 2 {\rm I}_3 
            \left( \sqrt{2} G_\mu  m_{\mathrm{Z}}^2 \right )^{1/2} \\
  g_+^{\gamma\mathrm{f}\bar{\mathrm{f}}} & =  & Q_{\mathrm{f}} 
    \left( 4 \pi \alpha( M_{\gamma^*}^2) \right )^{1/2} \\
  g_-^{\gamma\mathrm{f}\bar{\mathrm{f}}} & =  & 
        g_+^{\gamma\mathrm{f}\bar{\mathrm{f}}}
\end{eqnarray}

\noindent
where $Q_{\mathrm{f}}$ is the charge of the fermion $\mathrm{f}$ in units of 
the charge of the positron, and the electromagnetic coupling constant 
$\alpha( M_{\gamma^*}^2)$ is evaluated at the scale of the virtual photon 
mass $ M_{\gamma^*}^2$. ${\rm I}_3$ is the third component of the weak 
isospin ($\pm 1/2$), $\sin^2\!\overline{\theta}_{\mathrm{W}}$ is the effective 
value of the square of the sine of the Weinberg angle and $G_\mu$ is the value 
of the Fermi coupling constant.
The effective couplings to the Z absorb the relevant electroweak 
radiative corrections at the scale of the Z \cite{lep1_yellowreport}. They are 
obtained by the substitutions:
\begin{eqnarray}
    \sin^2\!\theta_{\mathrm{W}} & \rightarrow &  
       \sin^2\!\overline{\theta}_{\mathrm{W}} \\
    \frac{e^2}{4\sin^2\!\theta_{\mathrm{W}}\cos^2\!\theta_{\mathrm{W}}} & 
       \rightarrow &  \sqrt{2} G_\mu  m_{\mathrm{Z}}^2 
\end{eqnarray}

   The experimental signature of a 
$\mathrm{e^+ e^-} \rightarrow \mathrm{Z}\mathrm{Z}$ process is a final 
state with four fermions, due to the unstability of the $\mathrm{Z}$ particle. 
A distinctive feature is that the invariant masses of the two pairs,
$\mathrm{f}\bar{\mathrm{f}}$ and $\mathrm{f^\prime}\bar{\mathrm{f^\prime}}$, 
peak at the Z mass $ m_{\mathrm{Z}}$.  The angular distribution of the decay 
products keeps information on the average polarization of the 
$\mathrm{Z}$ boson. In addition, the $\mathrm{Z}$ decay amplitude has to 
be included for a correct treatment of spin correlations. 
Assuming that fermion masses are negligible compared with $ m_{\mathrm{Z}}$, 
this amplitude is given in the rest frame of the $\mathrm{Z}$ by:
\begin{eqnarray}
 M_{\mathrm{Z}_i\mathrm{f}\bar{\mathrm{f}}}
    (\lambda_{\mathrm{Z}_i},\lambda,\overline{\lambda}) & = & 
    g_\lambda^{Zff}~ M_{\mathrm{Z}_i}~\delta_{\lambda,-\overline{\lambda}}~~
\left[ \epsilon_{\mathrm{Z}_i} \left( v_1 - i \lambda v_2 \right) \right] \\ 
 & & \nonumber \\
v_1 & = & \left(0, \cos\theta_f \cos\phi_f, \cos\theta_f \sin\phi_f, 
      -\sin\theta_f \right) \\
v_2 & = & \left(0, -\sin\phi_f, \cos\phi_f, 0 \right) 
\end{eqnarray}

\noindent
where $\hat{p} = (\sin\theta_f\cos\phi_f,\sin\theta_f\sin\phi_f,\cos\theta_f)$ 
is the direction of the fermion momentum and $\lambda,\overline{\lambda}$ are 
the helicities of fermion and antifermion, respectively.

\section*{Anomalous couplings in the 
$\bbox{\mathrm{\lowercase{e}^+ \lowercase{e}^-} 
  \rightarrow \mathrm{Z}\mathrm{Z}}$ process}

  Anomalous gauge boson couplings lead to interactions of the type shown in 
Figure \ref{fig:zzanom}. The coupling $\mathrm{Z}\mathrm{Z} V$, with 
$V=\mathrm{Z}$ or $\gamma$, does not exist in the Standard Model at tree level.
Only two anomalous couplings are possible if the $\mathrm{Z}$ bosons in the 
final state are on-shell, due to Bose-Einstein symmetry. 
In principle, five more couplings should be considered if at least one of the 
$\mathrm{Z}$ bosons is off-shell. However, as discussed in Appendix \ref{app:a}, 
the new terms must be of higher dimensionality and are suppressed by orders of 
$ m_{\mathrm{Z}}\Gamma_{\mathrm{Z}}/\Lambda^2$, where $\Lambda$ denotes a 
scale related to new physics. We will concentrate on the most 
general expression of the anomalous vertex function at lowest order 
\cite{hagiwara_et_al}:
\begin{eqnarray}
\Gamma^{\alpha\beta\mu}_{\mathrm{Z}\mathrm{Z} V} = 
    \frac{s-m_V^2}{ m_{\mathrm{Z}}^2} \left\{ \right. &
     i f_4^V~[ (q_{\mathrm {Z_1}}+q_{\mathrm {Z_2}})^\alpha~g^{\mu\beta} + 
               (q_{\mathrm {Z_1}}+q_{\mathrm {Z_2}})^\beta~g^{\mu\alpha} ] 
   + i f_5^V~\epsilon^{\alpha\beta\mu\rho}~
          (q_{\mathrm {Z_1}}-q_{\mathrm {Z_2}})_\rho
   \left. \right\} 
\end{eqnarray}

A non-zero value of $f_4^V$ leads to a C-violating,CP-violating
process, while terms associated to $f_5^V$ are P-violating, CP-conserving.
    Using again the formalism followed in \cite{hagiwara_et_al} we 
obtain the explicit expressions for the anomalous contributions:
\begin{eqnarray}
   M_{AC}^{f_4^V}
(\sigma,\overline{\sigma},\lambda_{\mathrm {Z_1}},\lambda_{\mathrm {Z_2}}) = 
    & -i~e f_4^V g_\sigma^{Vee}~\frac{s}{ m_{\mathrm{Z}}^2}~
   \delta_{\sigma,-\overline{\sigma}} \left [ 
    \epsilon_{\mathrm {Z_1}}^{0*} 
    (\epsilon_{\mathrm {Z_2}}^{1*} + i\sigma \epsilon_{\mathrm {Z_2}}^{2*}) + 
    \epsilon_{\mathrm {Z_2}}^{0*} 
    (\epsilon_{\mathrm {Z_1}}^{1*} + i\sigma \epsilon_{\mathrm {Z_1}}^{2*}) 
          \right ] \\
   M_{AC}^{f_5^V}
(\sigma,\overline{\sigma},\lambda_{\mathrm {Z_1}},\lambda_{\mathrm {Z_2}}) = 
    & -i~e f_5^V g_\sigma^{Vee}~
            \frac{\sqrt{s}}{ m_{\mathrm{Z}}^2}~
   \delta_{\sigma,-\overline{\sigma}}~
    (\epsilon^{1\alpha\beta\rho} + i\sigma \epsilon^{2\alpha\beta\rho} )~
    \epsilon_{\mathrm {Z_1}\alpha}^{*}~\epsilon_{\mathrm {Z_2}\beta}^{*}~
      (q_{\mathrm {Z_1}\rho}-q_{\mathrm {Z_2}\rho})
\end{eqnarray}

\noindent
   Note that no $(s-m_V^2)$ factors are present in the final expressions. 
Compared to the SM amplitude all anomalous contributions increase with the 
center-of-mass energy of the collision.
We want to bring the attention to the fact that these anomalous 
$\mathrm{Z}\mathrm{Z}\gamma$ couplings are different from those considered 
in the $\mathrm{e^+ e^-} \rightarrow \mathrm{Z}~\gamma$ 
anomalous process \cite{hagiwara_et_al}. 
Therefore, not only the two anomalous $\mathrm{Z}\mathrm{Z}\mathrm{Z}$ 
couplings, but all four anomalous parameters remain essentially 
unconstrained at present.

Anomalous $\mathrm{Z}\mathrm{Z} V$ couplings manifest in three ways:

\begin{itemize}
  \item A change in the observed total cross section 
         $\mathrm{e^+ e^-} \rightarrow \mathrm{Z}\mathrm{Z}$.
  \item A modification of the angular distribution of the $\mathrm{Z}$.
  \item A change in the average polarization of the $\mathrm{Z}$ bosons.
\end{itemize}

\begin{figure}[htbp]
\begin{center}
    \includegraphics[width=8cm]{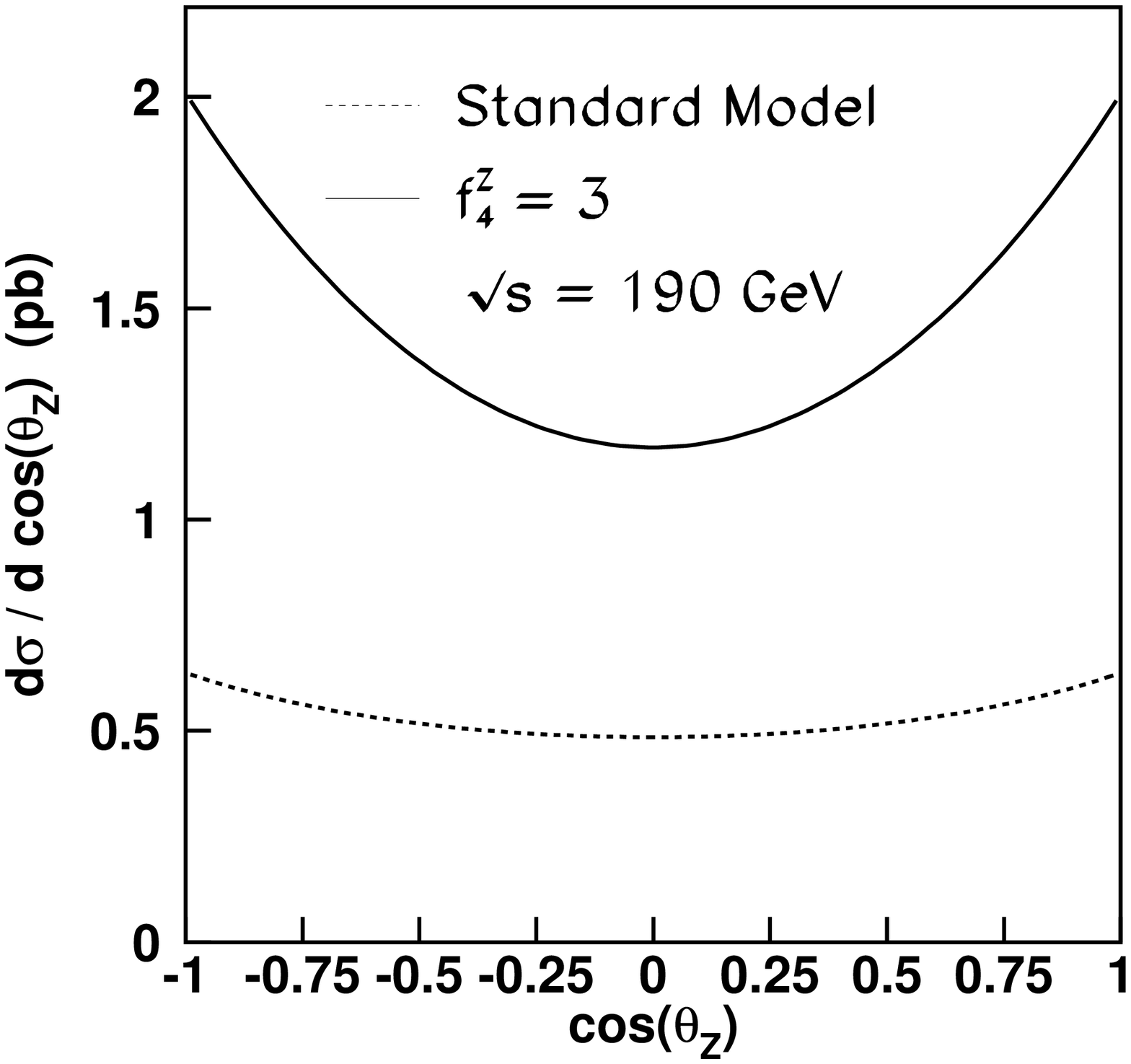}
    \includegraphics[width=8cm]{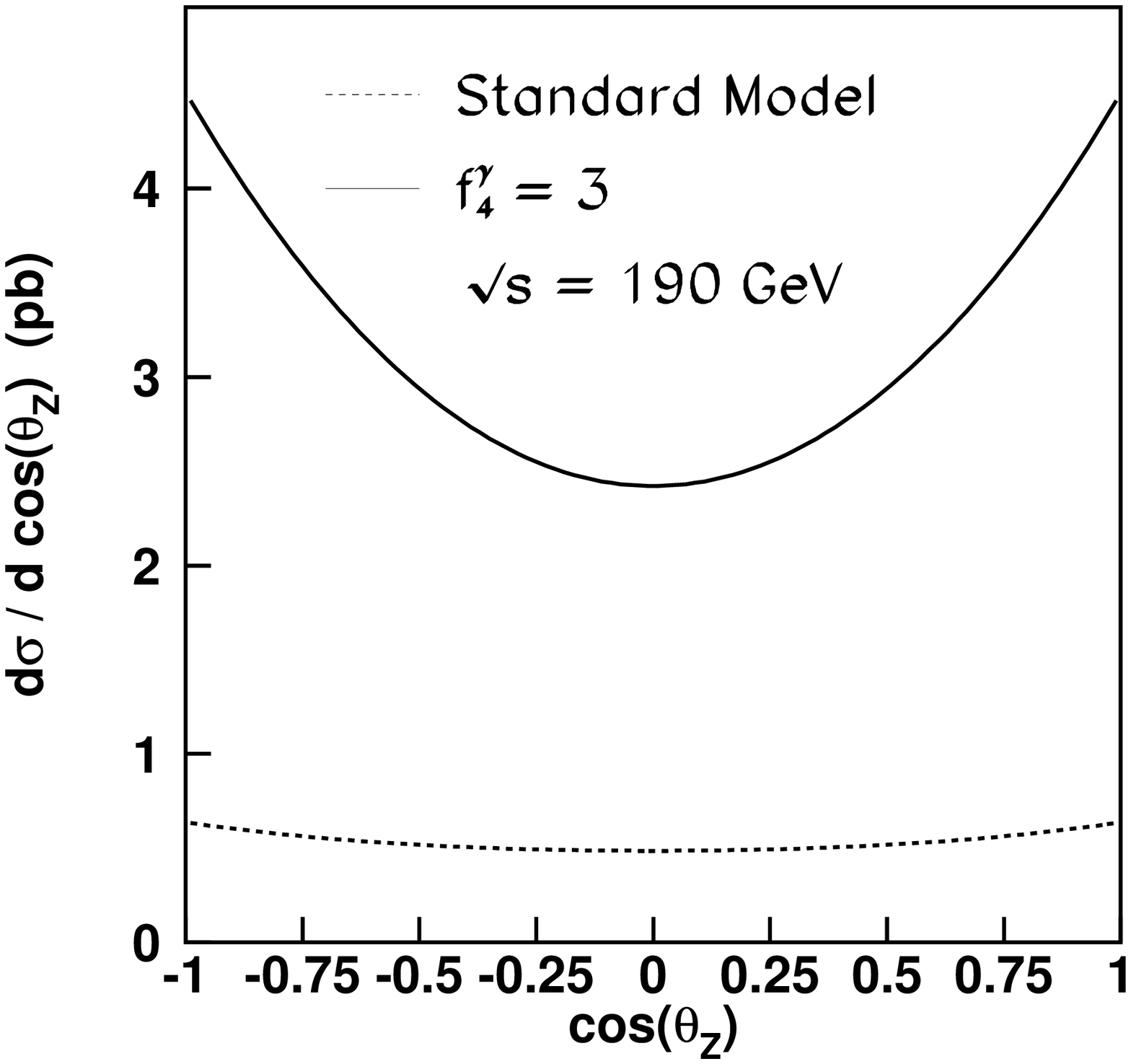} \\
    \includegraphics[width=8cm]{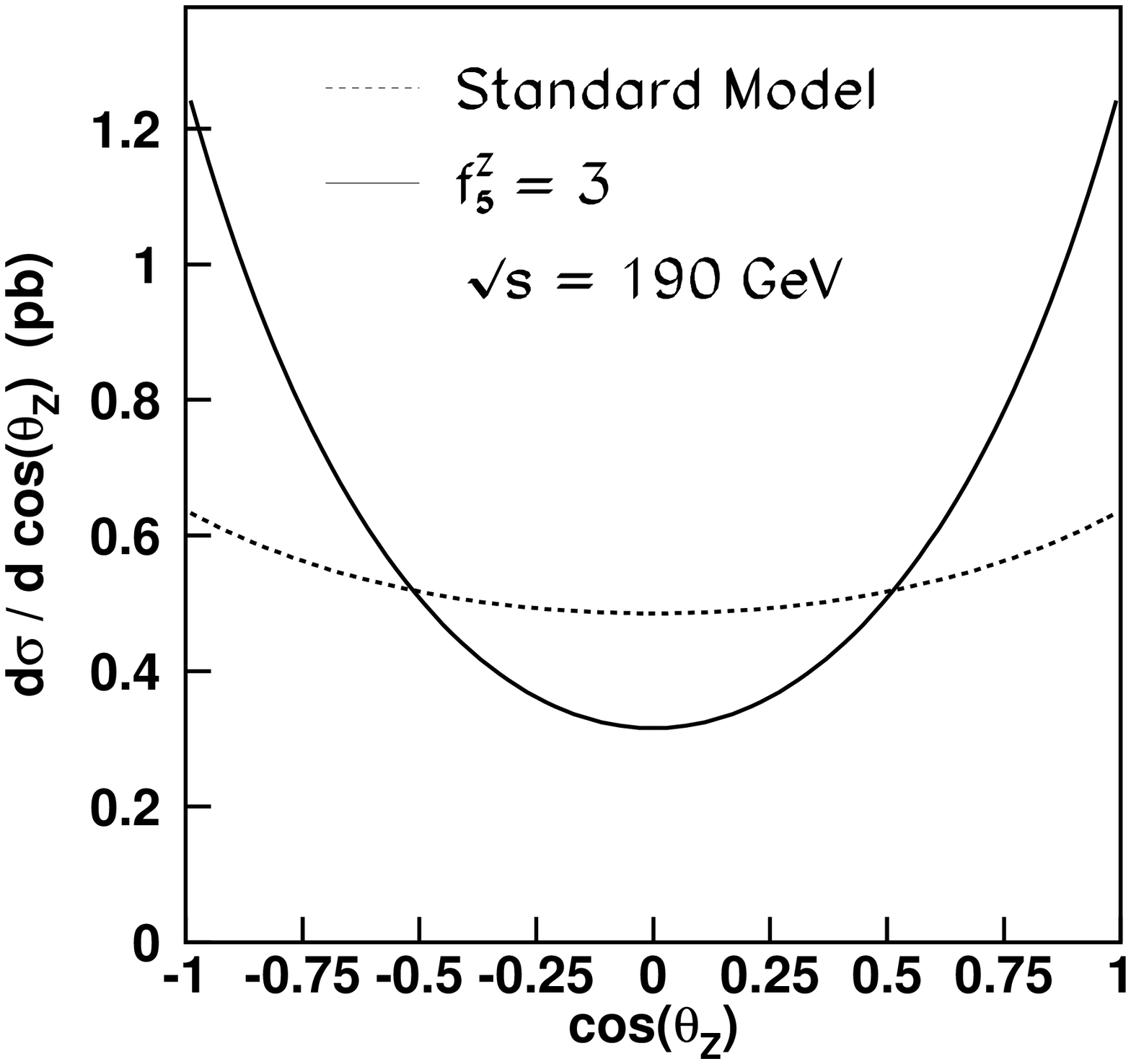}
    \includegraphics[width=8cm]{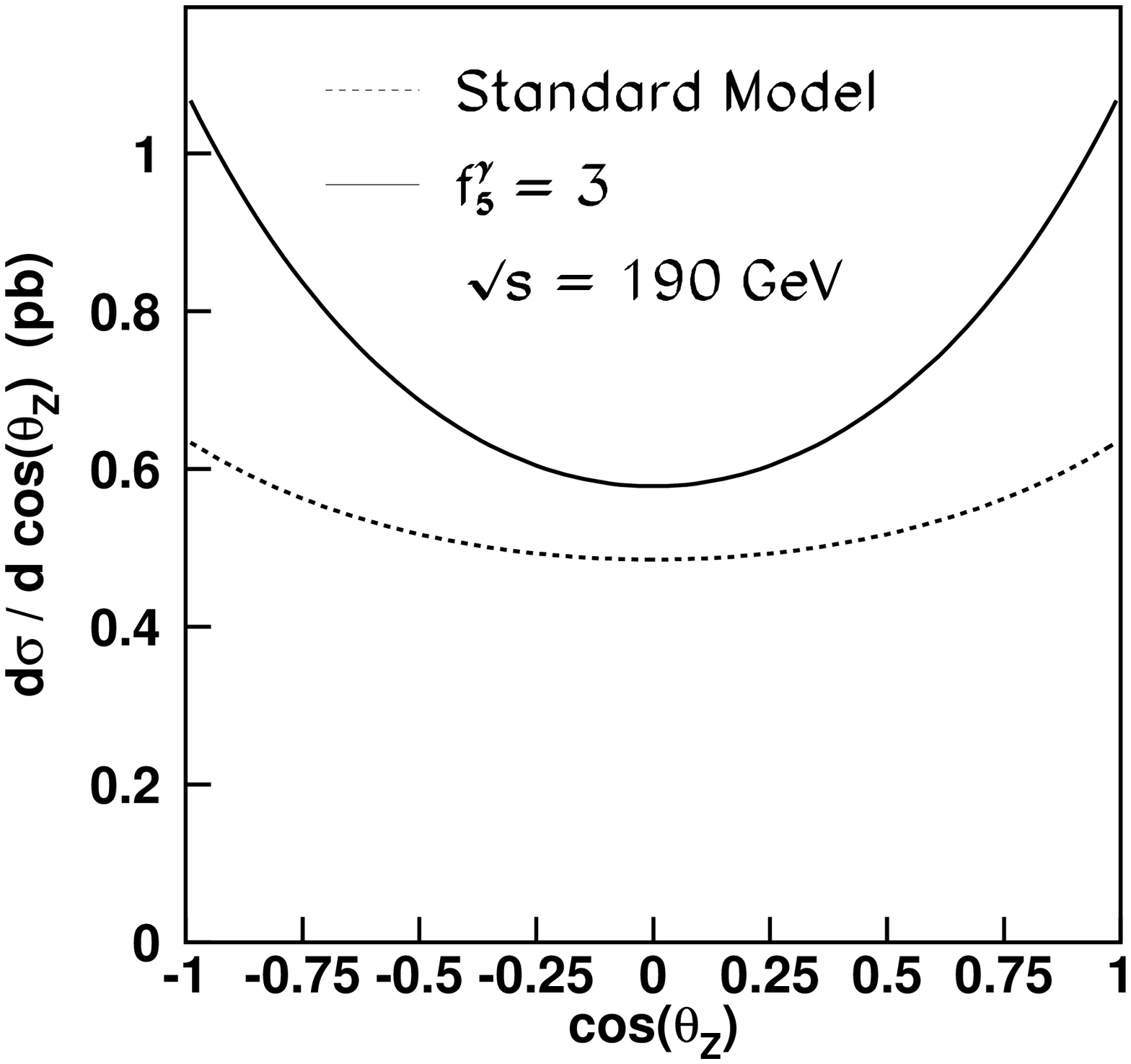}
\end{center}
\caption{ Effect of non-standard couplings in the 
          $\mathrm{e^+ e^-} \rightarrow \mathrm{Z}\mathrm{Z}$
         process at $\sqrt{s} = 190~\mathrm{\ Ge\kern -0.1em V}$. 
       A collision in the $\mathrm{e^+ e^-}$ center-of-mass system is assumed.
         The angle $\theta_Z$ is the polar angle of one of the 
         $\mathrm{Z}$ bosons and
         $d\sigma / d\cos\theta_Z$ is the differential cross section.}
\label{fig:angdis}
\end{figure}

\begin{figure}[htbp]
\begin{center}
    \includegraphics[width=8cm]{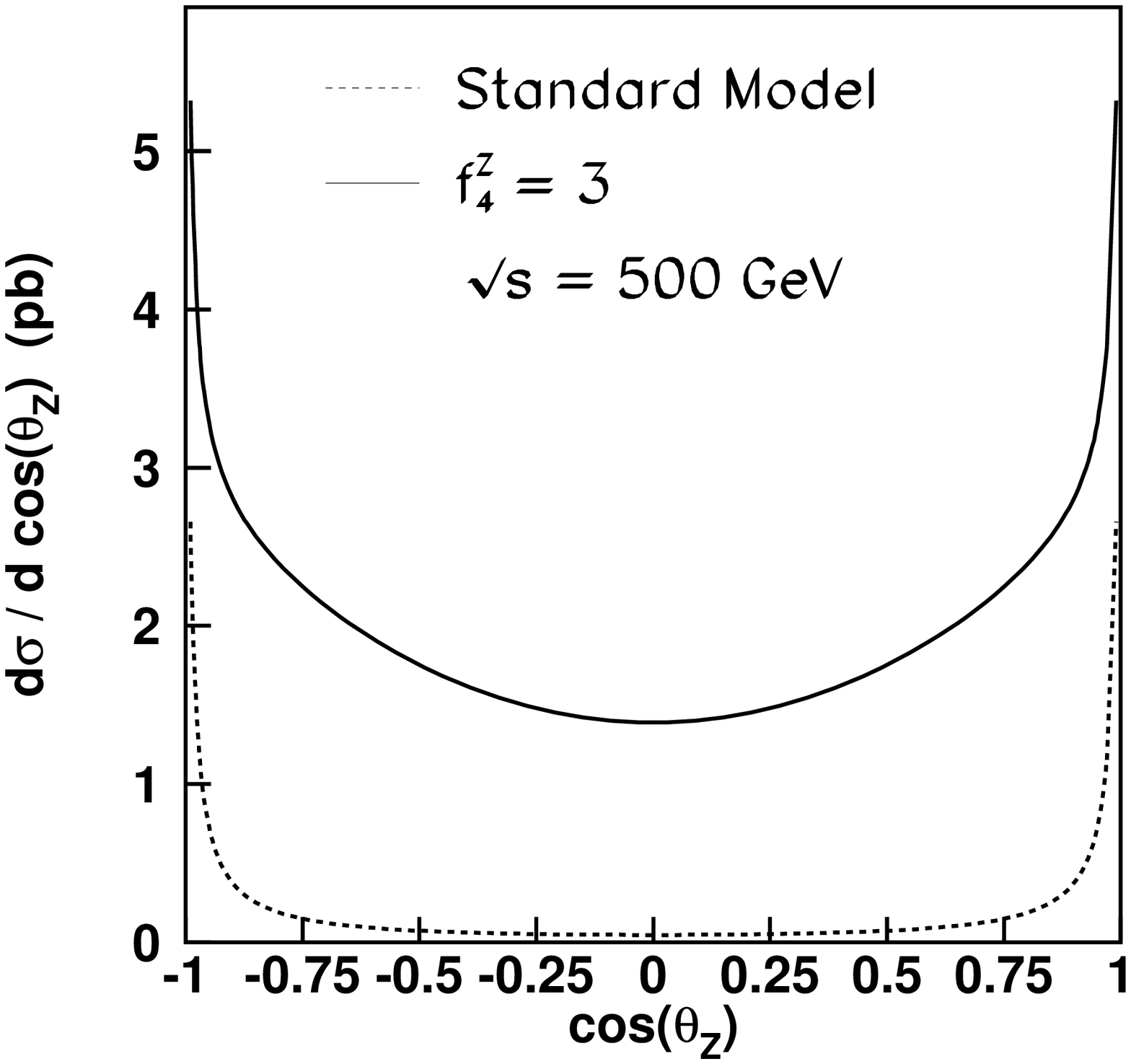}
    \includegraphics[width=8cm]{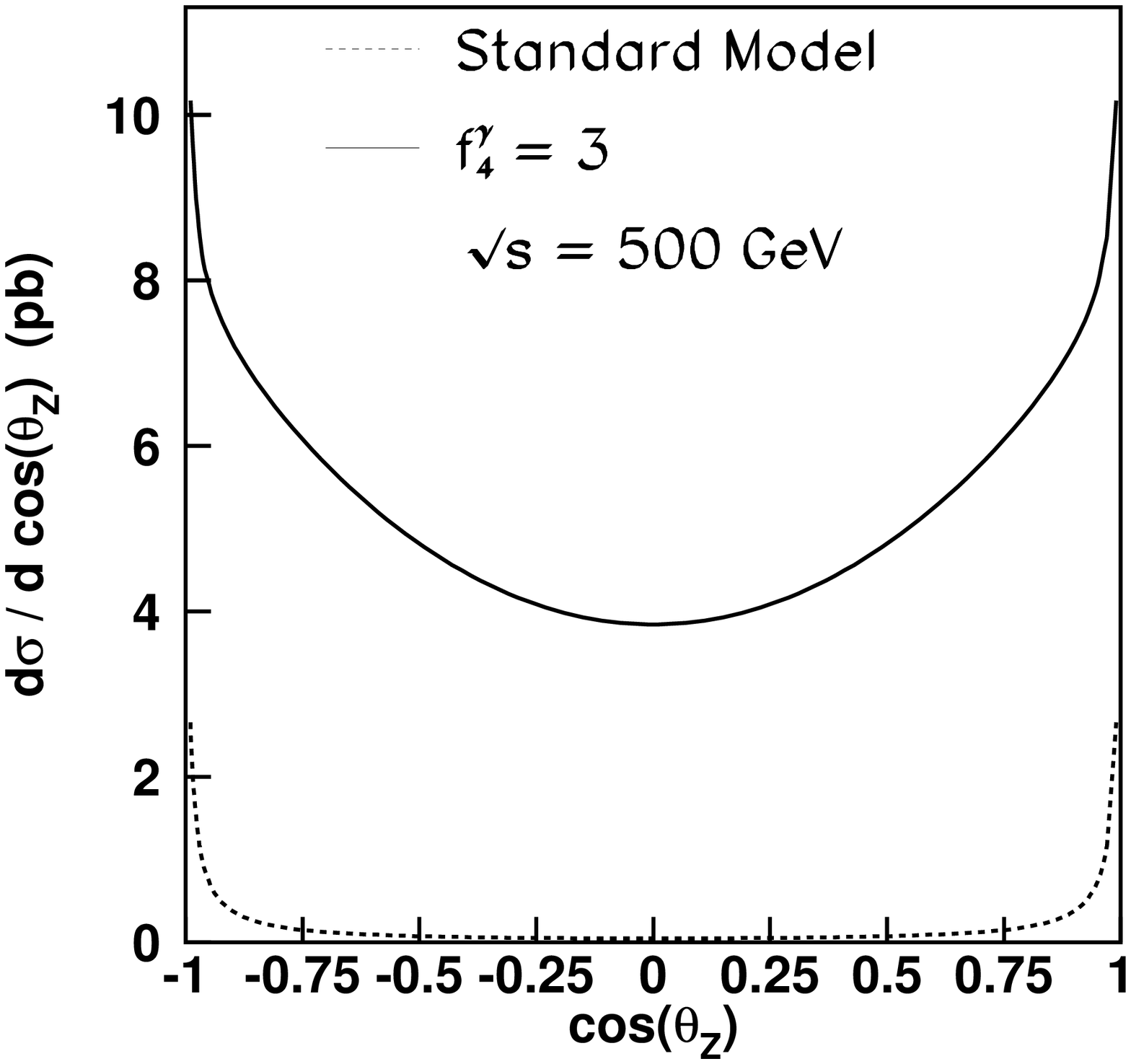} \\
    \includegraphics[width=8cm]{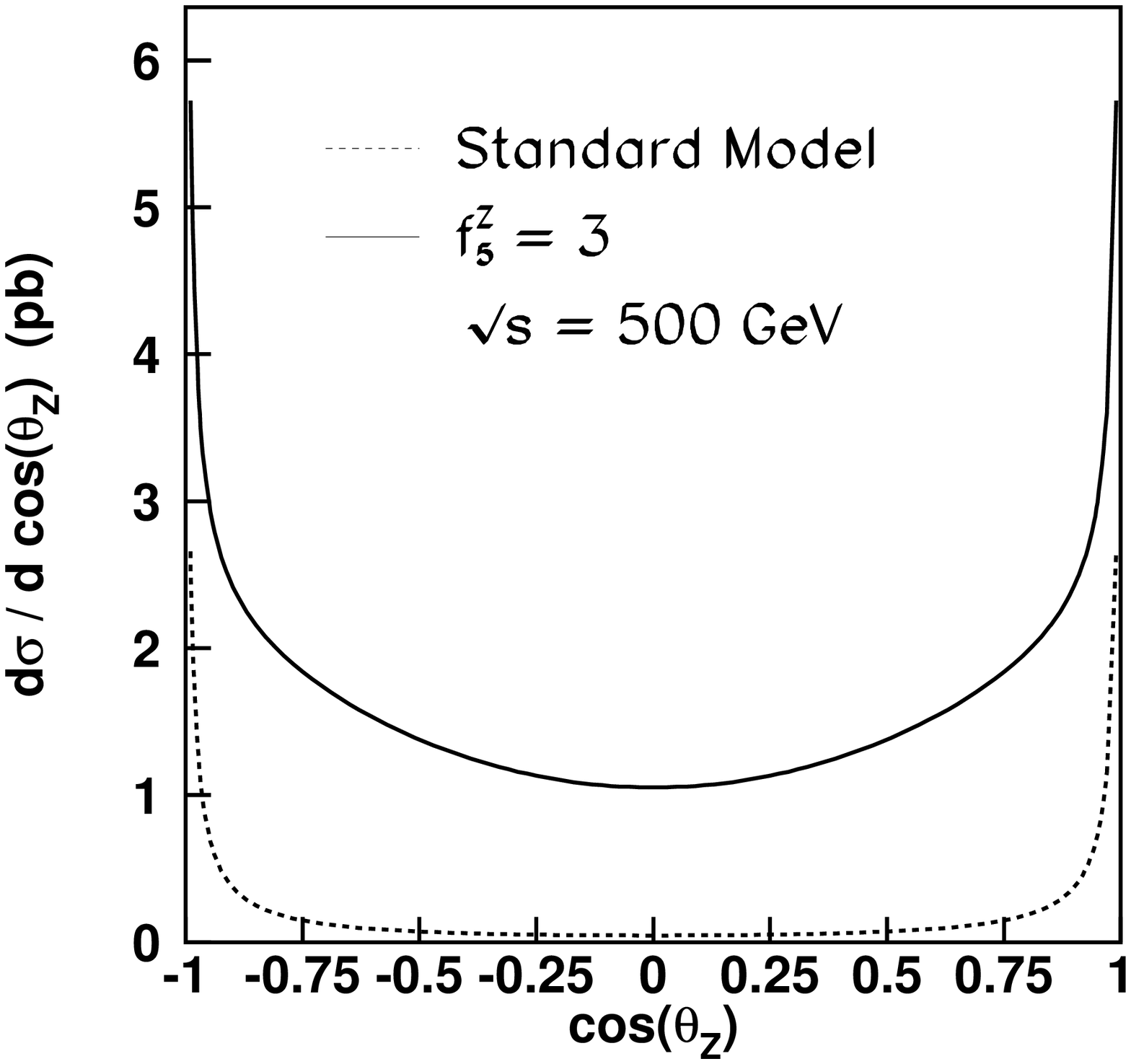}
    \includegraphics[width=8cm]{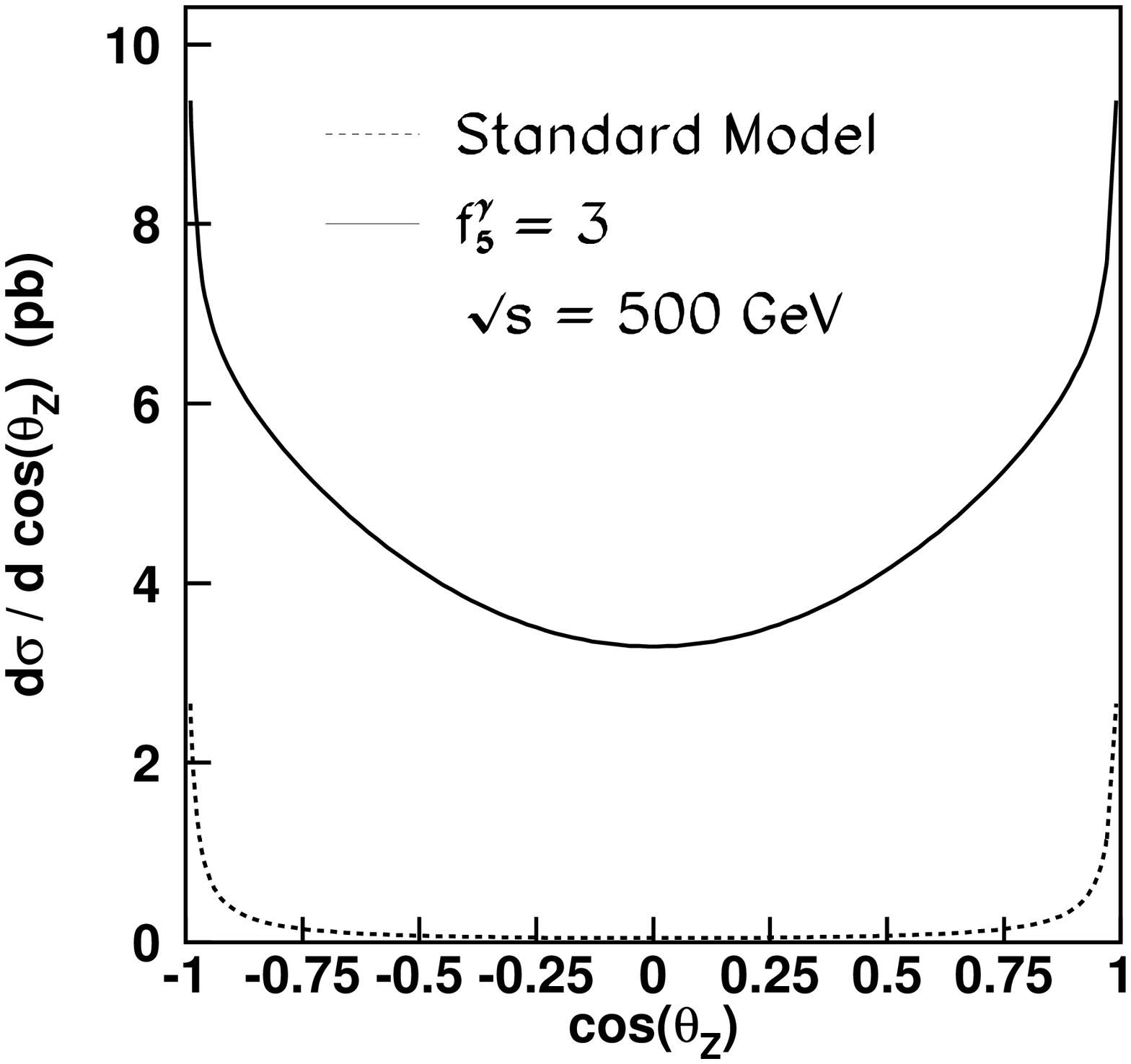}
\end{center}
\caption{ Effect of non-standard couplings in the 
          $\mathrm{e^+ e^-} \rightarrow \mathrm{Z}\mathrm{Z}$
         process at $\sqrt{s} = 500~\mathrm{\ Ge\kern -0.1em V}$. 
         A collision in the 
         $\mathrm{e^+ e^-}$ center-of-mass system is assumed.
         The angle $\theta_Z$ is the polar angle of one of the 
         $\mathrm{Z}$ bosons and
         $d\sigma / d\cos\theta_Z$ is the differential cross section.}
\label{fig:lc_angdis}
\end{figure}

 The effect of anomalous couplings in the 
$\mathrm{e^+ e^-} \rightarrow \mathrm{Z}\mathrm{Z}$ process 
at Born level is illustrated in 
Figures~\ref{fig:angdis} and~\ref{fig:lc_angdis} for 
the center-of-mass energies of $\sqrt{s}=190 \mathrm{\ Ge\kern -0.1em V}$ and 
$\sqrt{s}=500 \mathrm{\ Ge\kern -0.1em V}$, 
respectively. The anomalous distributions are 
determined for the values $f_i^V=3; i=4,5; V=\mathrm{Z},\gamma$. 
Both CP-violating and P-violating couplings are found to 
produce a global enhancement in the number of events. This increase is 
very clear at $\sqrt{s}=500 \mathrm{\ Ge\kern -0.1em V}$. There are moderate 
changes in the angular shape at $\sqrt{s}=190 \mathrm{\ Ge\kern -0.1em V}$.
At $\sqrt{s}=500 \mathrm{\ Ge\kern -0.1em V}$ the situation is different. 
The copious anomalous 
production at large polar angles starts to compensate the huge peaks of the 
SM differential cross section at low polar angles. The SM divergent behaviour 
happens in the limit $ m_{\mathrm{Z}} \ll \sqrt{s}$, where the process tends 
to a t-channel process with production of massless bosons.

\begin{figure}[htbp]
\begin{center}
    \includegraphics[width=8cm]{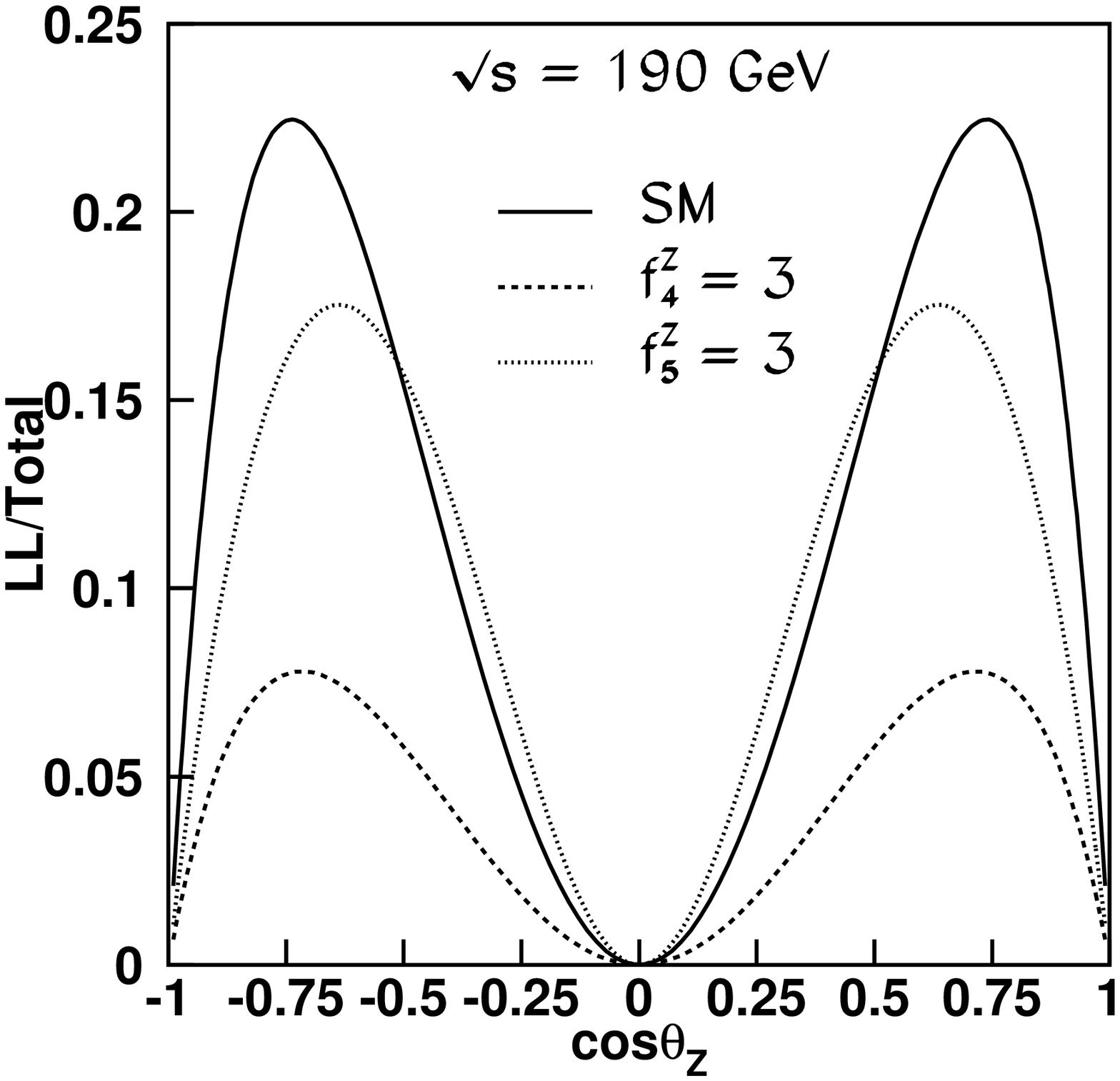}
    \includegraphics[width=8cm]{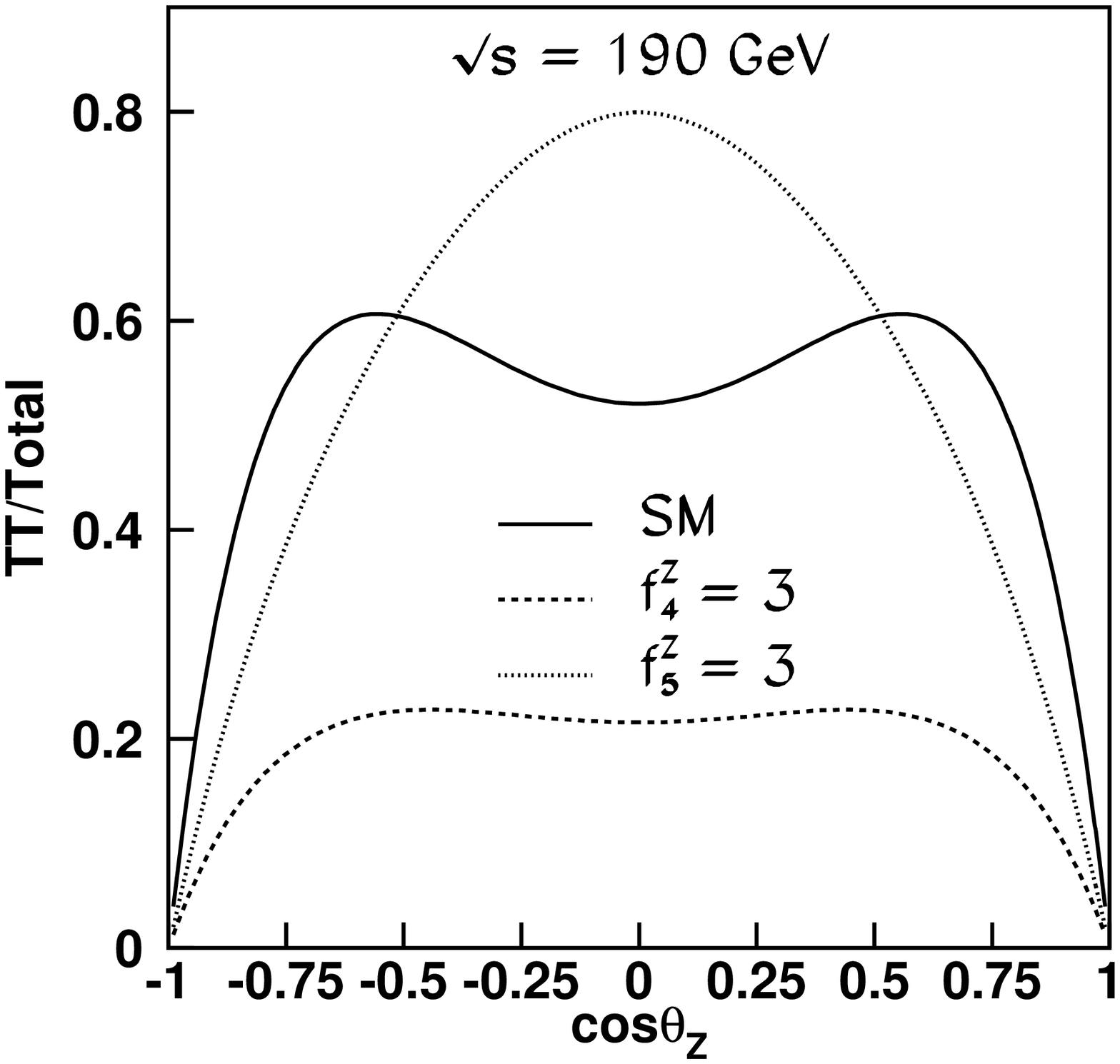}
    \includegraphics[width=8cm]{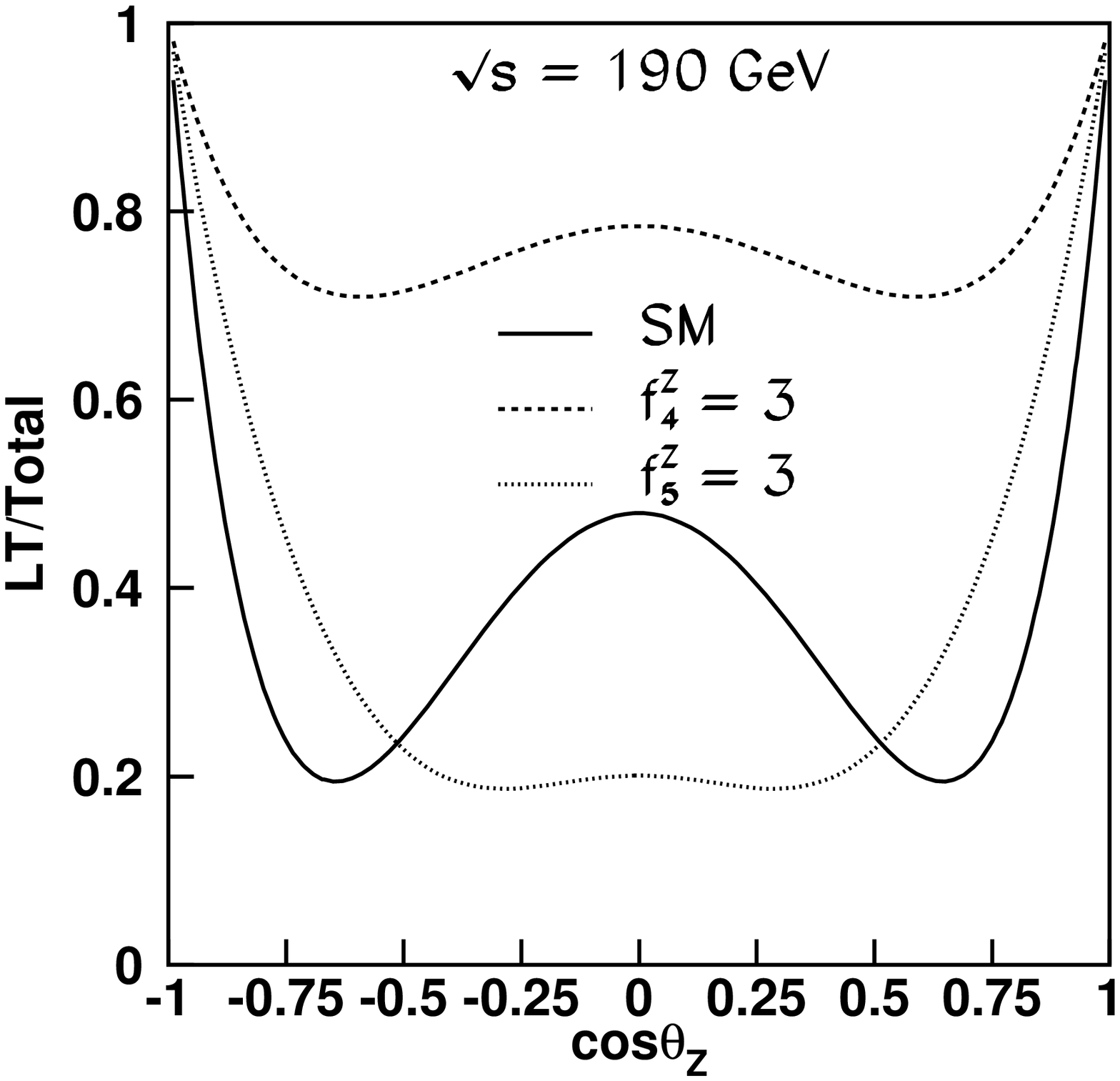}
\end{center}
\caption{ Proportion of events in which both Z are longitudinally polarized
          (top left),
          both Z are polarized transversely (top right) or one is 
          longitudinally and
          the other is transversely polarized (bottom) in a
          $\mathrm{e^+ e^-} \rightarrow \mathrm{Z}\mathrm{Z}$ process at 
          $\sqrt{s} = 190~\mathrm{\ Ge\kern -0.1em V}$. 
          Longitudinal and transverse polarizations are denoted by the labels
          ``L'' and ``T'', respectively.
          Predictions for SM
          (continuous line) and anomalous $\mathrm{Z}\mathrm{Z}\mathrm{Z}$ 
          couplings (dashed lines) are shown. A collision in the 
          $\mathrm{e^+ e^-}$ center-of-mass system is assumed.
          The angle $\theta_Z$ is the polar angle of one of the 
          $\mathrm{Z}$ bosons.}
\label{fig:polz}
\end{figure}

\begin{figure}[htbp]
\begin{center}
    \includegraphics[width=8cm]{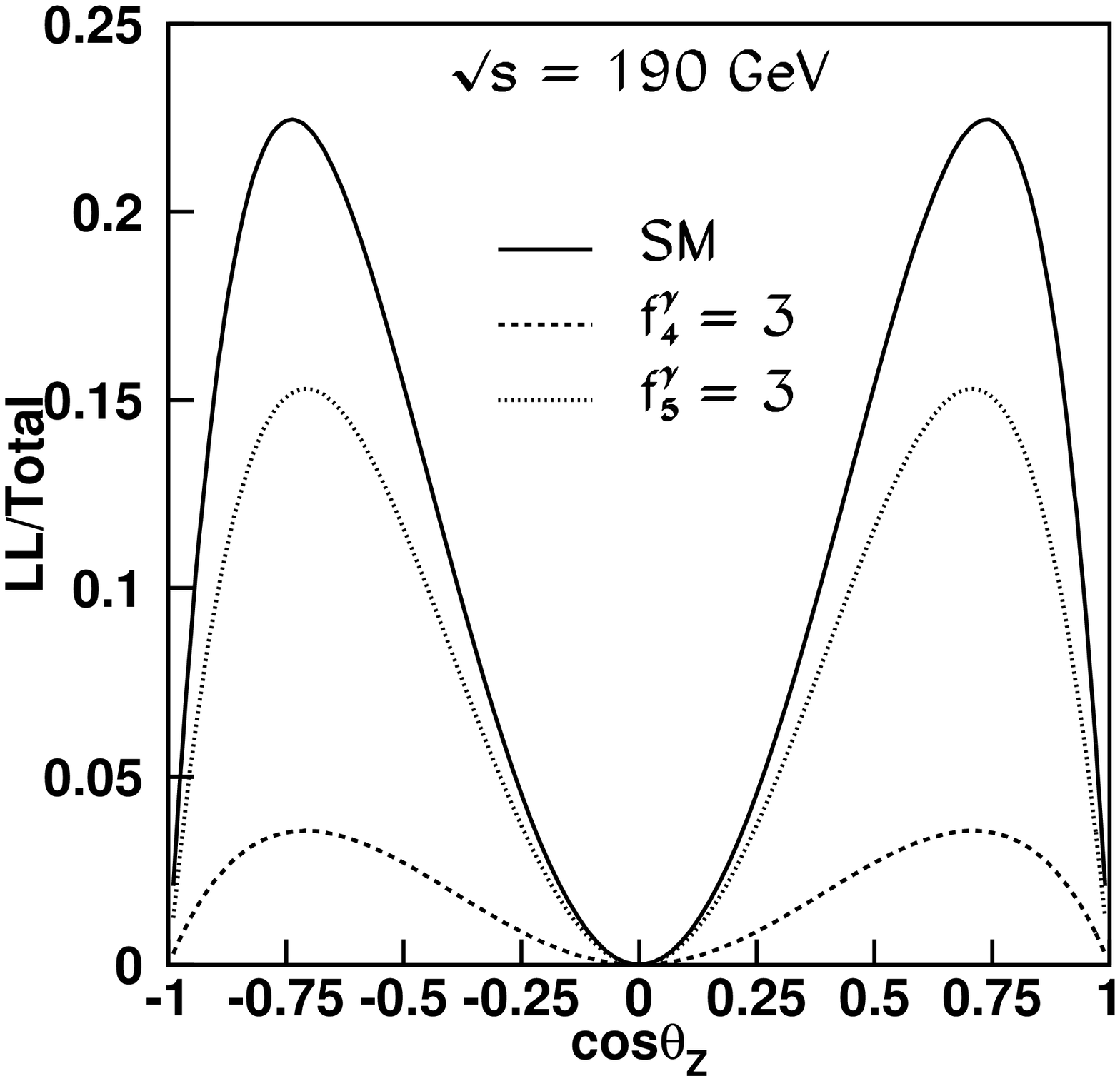}
    \includegraphics[width=8cm]{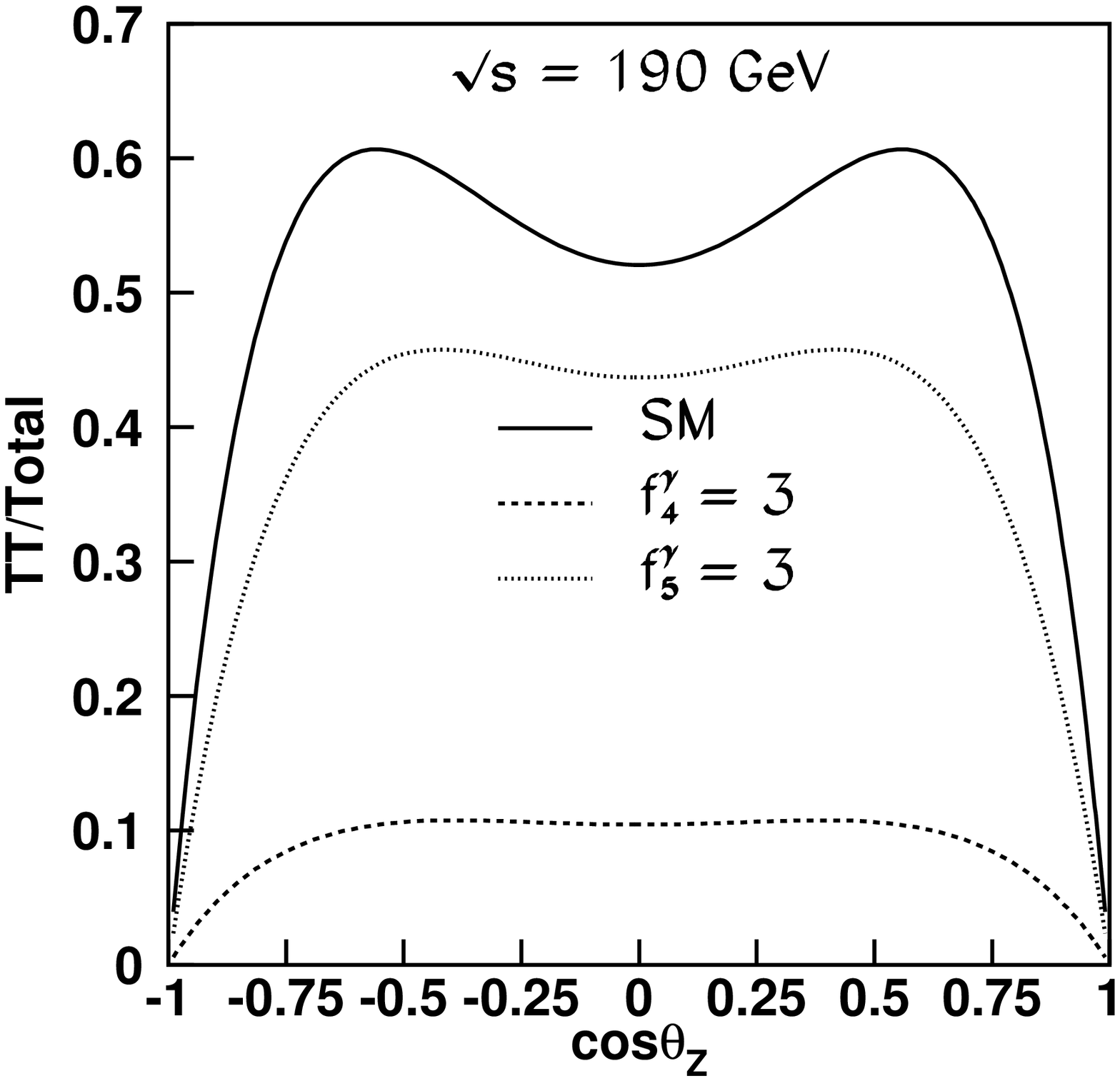}
    \includegraphics[width=8cm]{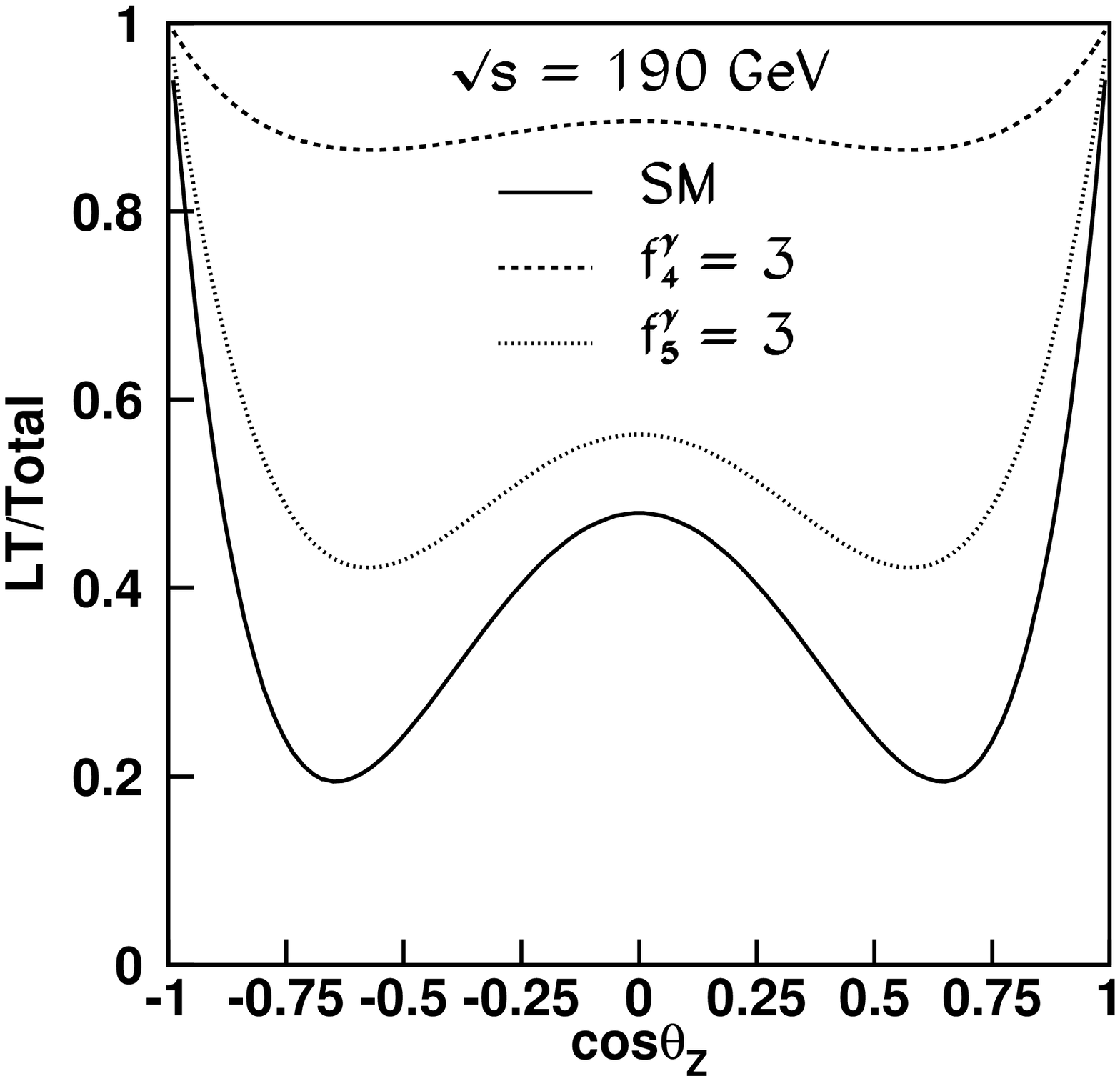}
\end{center}
\caption{ Proportion of events in which both Z are longitudinally polarized
          (top left), both Z are polarized transversely (top right) or one 
          is longitudinally and the other is transversely polarized (bottom) 
          in a $\mathrm{e^+ e^-} \rightarrow \mathrm{Z}\mathrm{Z}$ process at 
          $\sqrt{s} = 190~\mathrm{\ Ge\kern -0.1em V}$. 
          Longitudinal and transverse polarizations are denoted by the labels
          ``L'' and ``T'', respectively.
          Predictions for SM (continuous line) and anomalous 
          $\mathrm{Z}\mathrm{Z}\gamma$ couplings (dashed lines) are shown.
          A collision in the $\mathrm{e^+ e^-}$ 
          center-of-mass system is assumed.
          The angle $\theta_Z$ is the polar angle of one of the 
          $\mathrm{Z}$ bosons.}
\label{fig:polg}
\end{figure}

\begin{figure}[htbp]
\begin{center}
    \includegraphics[width=8cm]{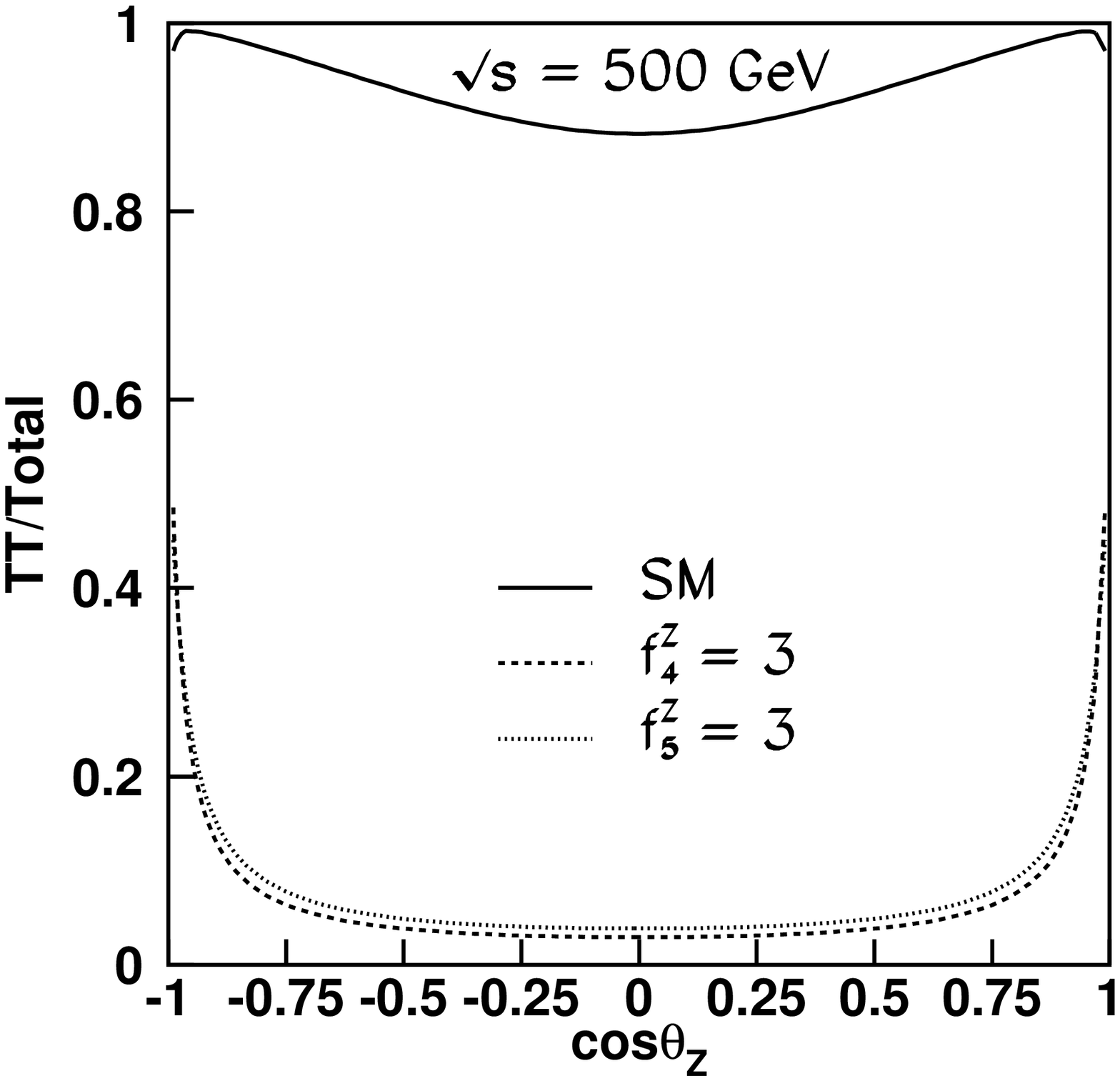}
    \includegraphics[width=8cm]{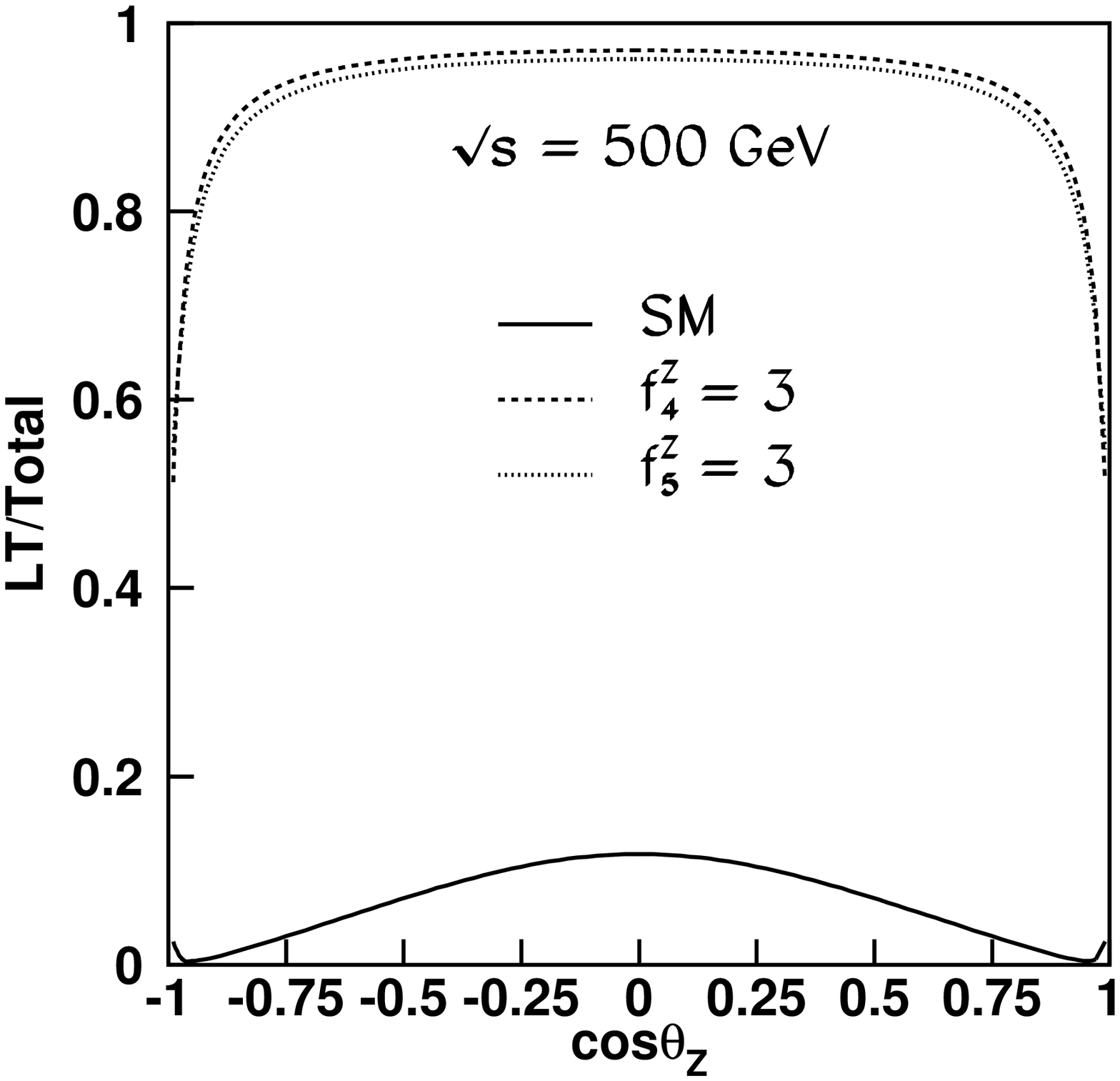} \\
    \includegraphics[width=8cm]{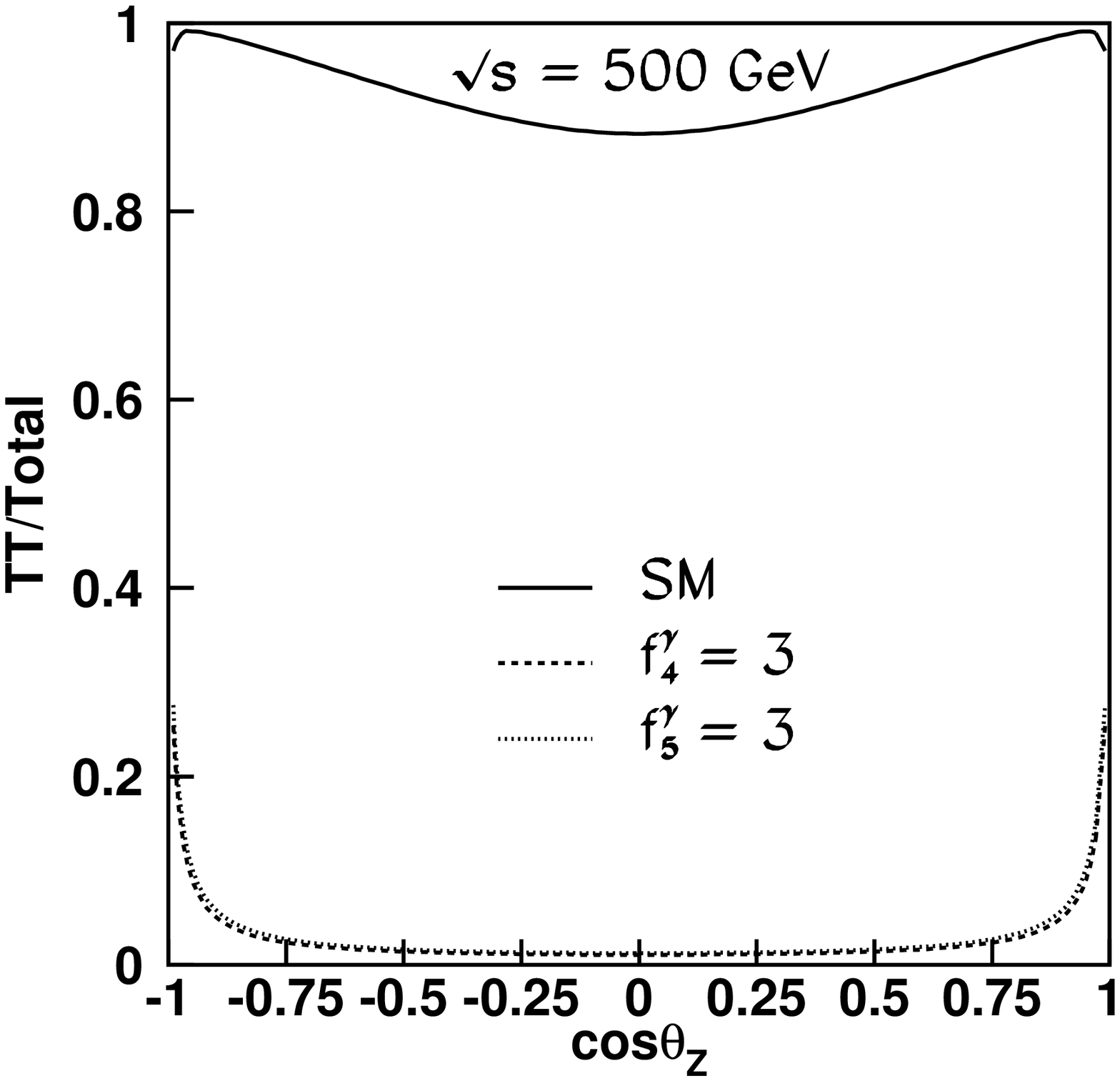}
    \includegraphics[width=8cm]{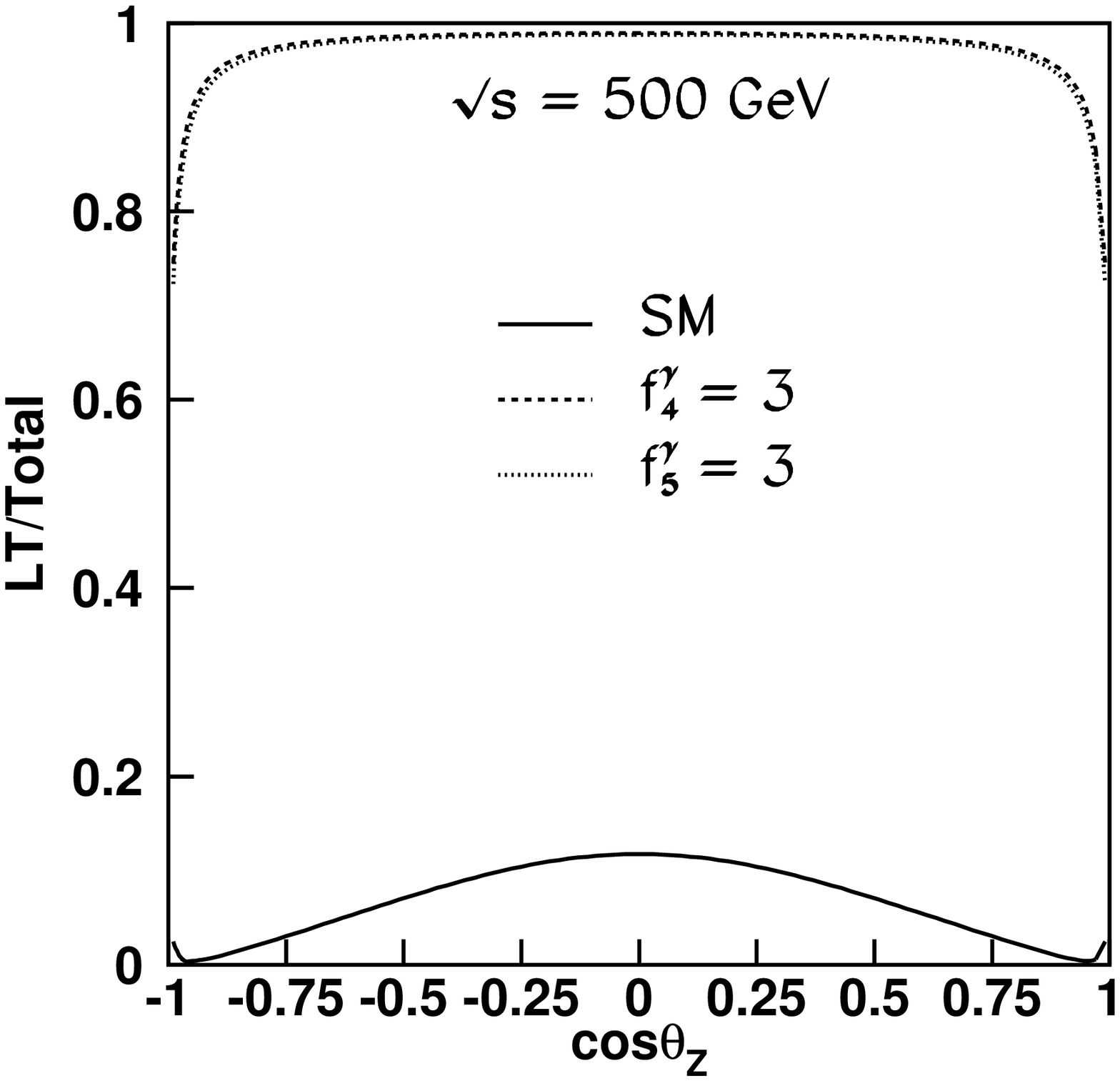}
\end{center}
\caption{ Proportion of events in which both Z are transversally polarized
          (left), or one is longitudinally and
          the other is transversely polarized (right) in a
          $\mathrm{e^+ e^-} \rightarrow \mathrm{Z}\mathrm{Z}$ process at 
          $\sqrt{s} = 500~\mathrm{\ Ge\kern -0.1em V}$. 
          Longitudinal and transverse polarizations are denoted by the labels
          ``L'' and ``T'', respectively.
          The two figures
          at the top (bottom) illustrate the effect of possible anomalous 
          $\mathrm{Z}\mathrm{Z}\mathrm{Z}$ ($\mathrm{Z}\mathrm{Z}\gamma$) 
          couplings. The Standard Model predictions are 
          shown with a continuous line, whereas the anomalous predictions 
          are shown with dashed lines. A collision in the 
          $\mathrm{e^+ e^-}$ center-of-mass 
          system is assumed. The angle $\theta_Z$ is the polar angle of one of 
          the $\mathrm{Z}$ bosons. The case in which both Z are longitudinally 
          polarized is not shown. It accounts for less than $0.5\%$ of the 
          total.}
\label{fig:lc_pol}
\end{figure}

   Anomalous couplings also modify the average polarizations of the
$\mathrm{Z}$ bosons, as shown in Figures~\ref{fig:polz},~\ref{fig:polg} 
and~\ref{fig:lc_pol}. At $\sqrt{s}=190 \mathrm{\ Ge\kern -0.1em V}$ the 
observed change depends on the
particular size and type of the anomalous coupling under consideration. For 
CP-violating couplings there is always an increase of the production of bosons 
with different polarizations (longitudinal versus transverse).
 At $\sqrt{s}=500 \mathrm{\ Ge\kern -0.1em V}$ all couplings show a similar 
behaviour: an enhancement of longitudinal-transverse production 
and a suppression of transverse-transverse production.
This is an interesting feature, since the SM process has the opposite 
behaviour. The fraction of states in which the two bosons are longitudinally 
polarized is below $0.5\%$ at these energies, both for SM and for anomalous 
production. What is physically observable is a modification of the angular 
distributions in the center-of-mass frame of the 
$\mathrm{Z} \rightarrow \mathrm{f}\bar{\mathrm{f}}$ decays: in 
absence of anomalous couplings, both $\mathrm{Z}$ decays will proceed 
preferentially along the direction of the $\mathrm{Z}$ momenta, whereas one 
of the decays will preferentially occur at $90^\circ$ if the process is 
highly anomalous.

   Summarizing, at energies close to the threshold of 
$\mathrm{Z}\mathrm{Z}$ production the 
sensitivity to anomalous couplings is weak. For an integrated luminosity 
of 200 \mbox{pb$^{-1}$}, tens of events are expected to be selected. 
The main anomalous effect is an increase in the cross section, and one expects 
small improvements from the variations in the angular distributions. At higher 
energies, with luminosities of the order of 100 \mbox{fb$^{-1}$}, all 
anomalous effects 
contribute coherently to enhance the sensitivity: a huge increase in the cross 
section, especially at large polar angles, and a clear correlation between the 
angular distributions of the $\mathrm{Z}$ decay products.

\section*{Inclusion of anomalous couplings. Reweighting procedure} 

     In order to take into account anomalous effects with sufficient accuracy, 
the correct matrix element structure has to be implemented. In many cases, 
and from the practical point of view, the generation of 
events for different values of anomalous couplings is not convenient. 
A more attractive method is to set up a procedure to obtain the 
Monte Carlo anomalous distributions as a function of a single set 
of generated events. This is the role of reweighting methods.

     Let us consider a set of events generated according to the Standard Model 
differential cross section. New distributions taking into account the anomalous
couplings are obtained when every event is reweighted by the factor:
\begin{eqnarray}
   W^{\mathrm{Z}\mathrm{Z}}(\sigma,\lambda,\lambda^\prime;\bar{\Omega}) & 
     \equiv & \frac{\left | {\displaystyle \sum_{\lambda_{\mathrm {Z_1}},
   \lambda_{\mathrm {Z_2}}} (M_{\mathrm{Z}\mathrm{Z}}+M_{AC})~
                 M_{\mathrm {Z_1}\mathrm{f}\bar{\mathrm{f}}}~
   M_{\mathrm {Z_2}\mathrm{f^\prime}\bar{\mathrm{f^\prime}}} } \right |^2}
        {\left | {\displaystyle \sum_{\lambda_{\mathrm {Z_1}},
   \lambda_{\mathrm {Z_2}}} M_{\mathrm{Z}\mathrm{Z}}~
                 M_{\mathrm {Z_1}\mathrm{f}\bar{\mathrm{f}}}~
  M_{\mathrm {Z_2}\mathrm{f^\prime}\bar{\mathrm{f^\prime}}} } \right |^2}
\end{eqnarray}

   The weight 
$W^{\mathrm{Z}\mathrm{Z}}(\sigma,\lambda,\lambda^\prime;\bar{\Omega})$ 
depends on the helicities of the initial electron ($\sigma$) and of the final 
fermions ($\lambda$,$\lambda^\prime$).  It also depends on the kinematic 
variables defining the phase space ($\bar{\Omega}$). For convenience we 
choose the following set:

\begin{itemize}
\item The invariant masses of the $\mathrm{f}\bar{\mathrm{f}}$ and 
      $\mathrm{f^\prime}\bar{\mathrm{f^\prime}}$ systems: 
      $ M_{\mathrm {Z_1}}$, $ M_{\mathrm {Z_2}}$.
\item The polar and azimuthal angles of the $\mathrm{f}\bar{\mathrm{f}}$ 
          system : 
      $\theta_{\mathrm{Z}}$, $\phi_{\mathrm{Z}}$.
\item The polar and azimuthal angles of the fermion $\mathrm{f}$ 
       after a Lorentz boost 
      to the rest frame of the $\mathrm{f}\bar{\mathrm{f}}$ system: 
           $\theta_f$, $\phi_f$.
\item The polar and azimuthal angles of the fermion $\mathrm{f}^\prime$ 
       after a 
      Lorentz boost to the rest frame of the 
        $\mathrm{f^\prime}\bar{\mathrm{f^\prime}}$ system: 
      $\theta_{f^\prime}$, $\phi_{f^\prime}$.
\end{itemize}

   The previous result can be extended to take into account other non-resonant 
diagrams like $\mathrm{e^+ e^-} \rightarrow \mathrm{Z} \gamma^* 
\rightarrow \mathrm{f}\bar{\mathrm{f}} 
\mathrm{f^\prime}\bar{\mathrm{f^\prime}}$ and
$\mathrm{e^+ e^-} \rightarrow \gamma^* \gamma^* 
\rightarrow \mathrm{f}\bar{\mathrm{f}} 
\mathrm{f^\prime}\bar{\mathrm{f^\prime}}$. These 
diagrams can not be neglected for a correct analysis of double Z resonant 
production~\cite{bardin}. The final weight is:
\begin{eqnarray}
   W^{\mathrm{V}\mathrm{V}}(\sigma,\lambda,\lambda^\prime;\bar{\Omega}) \equiv 
   \left | 1 + \frac{\displaystyle {\cal D}_{\mathrm {Z_1}}( M_{\mathrm {Z_1}}^2) 
                 {\cal D}_{\mathrm {Z_2}}( M_{\mathrm {Z_2}}^2)
              \sum_{\lambda_{\mathrm {Z_1}},\lambda_{\mathrm {Z_2}}} M_{AC}~
                 M_{\mathrm {Z_1}\mathrm{f}\bar{\mathrm{f}}}~
          M_{\mathrm {Z_2}\mathrm{f^\prime}\bar{\mathrm{f^\prime}}} }
                 {\displaystyle 
     \sum_{\mathrm {V_\circ},\mathrm {V_\circ^\prime}} 
       {\cal D}_{\mathrm {V_\circ}}( M_{\mathrm {Z_1}}^2) 
       {\cal D}_{\mathrm {V_\circ^\prime}}( M_{\mathrm {Z_2}}^2) 
                            \sum_{\lambda_{\mathrm {V_\circ}},i
         \lambda_{\mathrm {V_\circ^\prime}}} 
                 M_{\mathrm {V_\circ}\mathrm {V_\circ^\prime}}~
                 M_{\mathrm {V_\circ}\mathrm{f}\bar{\mathrm{f}}}~
     M_{\mathrm {V_\circ^\prime}\mathrm{f^\prime}\bar{\mathrm{f^\prime}}} } 
                        \right |^2
\end{eqnarray}

\noindent
where a sum on all intermediate 
$\mathrm {V_\circ} \in \{\mathrm{Z},\gamma^*\}$ is assumed. 
The propagator factors ${\cal D}_{\mathrm {V_\circ}}$ are defined as follows:
\begin{eqnarray}
   {\cal D}_{\mathrm {Z}}(q^2) & = & 
 \frac{1}{q^2- m_{\mathrm{Z}}^2 + i~\Gamma_{\mathrm{Z}}~q^2/ m_{\mathrm{Z}}} \\
   {\cal D}_\gamma(q^2) & = & \frac{1}{q^2}
\end{eqnarray}

\noindent
where the imaginary component takes into account the energy 
dependence of the $\mathrm{Z}$ width around the resonance. 
The expressions for $M_{\mathrm{Z}\gamma^*}$, 
$M_{\gamma^*\gamma^*}$, $M_{\gamma^*\mathrm{f}\bar{\mathrm{f}}}$ 
are obtained by the same method used for 
$M_{\mathrm{Z}\mathrm{Z}}$ and $M_{\mathrm{Z}\mathrm{f}\bar{\mathrm{f}}}$. 
Explicitly, they can be obtained by substituting $\mathrm{Z}$ by $\gamma^*$ 
where necessary:
\begin{eqnarray}
    \epsilon_{\mathrm{Z}}(\lambda_{\mathrm{Z}}) & \rightarrow &  
    \epsilon_{\gamma^*}(\lambda_\gamma) \\
     M_{\mathrm {Z_1}} & \rightarrow &  M_{\gamma^*} \\
    g_+^{\mathrm{Z}\mathrm{f}\bar{\mathrm{f}}} & 
      \rightarrow & g_+^{\gamma\mathrm{f}\bar{\mathrm{f}}} \\
    g_-^{\mathrm{Z}\mathrm{f}\bar{\mathrm{f}}} & \rightarrow & 
            g_-^{\gamma\mathrm{f}\bar{\mathrm{f}}}
\end{eqnarray}

   Weights according to 
$W^{\mathrm{V}\mathrm{V}}(\sigma,\lambda,\lambda^\prime;\bar{\Omega})$ have 
been implemented in a FORTRAN program. The approach is well suited for events 
generated with the PYTHIA 
$\mathrm{e^+ e^-} \rightarrow \mathrm{Z}/\gamma^*~\mathrm{Z}/\gamma^* 
\rightarrow \mathrm{f}\bar{\mathrm{f}} 
\mathrm{f^\prime}\bar{\mathrm{f^\prime}}$ 
generator \cite{pythia}. This implementation will be identified as 
the ``NC08 approach'', because all eight neutral conversion diagrams are 
considered. Several checks have been done in order to make
sure that the calculations are correct. There is also agreement with the 
results obtained in \cite{jadach} and in \cite{biebel_and_others}.

   After reweighting, distributions according to given 
values of the anomalous couplings $f_4^V,f_5^V$ are obtained. 
Detector effects are correctly taken into account if events are 
reweighted at generator level.

\section*{Initial state radiation effects}

    There are several references providing valuable information on the 
$\mathrm{e^+ e^-} \rightarrow \mathrm{Z}\mathrm{Z}$ process. 
A specific SM generator for 
$\mathrm{e^+ e^-} \rightarrow \mathrm{Z}/\gamma^*~\mathrm{Z}/\gamma^*~(\gamma) 
\rightarrow \mathrm{f}\bar{\mathrm{f}} 
\mathrm{f^\prime}\bar{\mathrm{f^\prime}} (\gamma)$
without anomalous couplings exists in PYTHIA \cite{pythia}. The calculation 
reported in the previous section is well suited for this MC generator, but
initial state radiation effects (ISR) need to be taken into account. We assume
that the differential cross section can be expressed as follows:
\begin{eqnarray}
\frac{d \sigma(s)}{d ({\rm Phase~Space})} = \int ds^\prime R(s,s^\prime) ~
         \frac{d \sigma(s^\prime)}{d ({\rm Phase~Space^\prime})} 
           \label{eq:radiator}
\end{eqnarray}

\noindent
where $\sigma(s^\prime)$ is the (undressed) cross 
section evaluated at a scale $s^\prime$, and $\sigma(s)$ is the cross 
section after inclusion of 
ISR effects. The radiator factor $R(s,s^\prime)$ is 
a ``universal'' radiator, that is, independent of specific details of the 
matrix element. With this assumption ISR effects are accounted for by 
evaluating the matrix element in the center-of-mass system of 
the four fermions, at the corresponding scale $s^\prime$.

  In Reference \cite{jadach} a specific
$\mathrm{e^+ e^-} \rightarrow \mathrm{Z} \mathrm{Z} (\gamma) 
\rightarrow \mathrm{f}\bar{\mathrm{f}} 
\mathrm{f^\prime}\bar{\mathrm{f^\prime}} (\gamma)$ 
generator for anomalous couplings studies is presented. 
It takes into account ISR effects 
with the YFS approach \cite{yfs} up to ${\cal O}(\alpha^2)$ leading-log. 
It has some limitations, like the absence of conversion diagrams mediated by 
virtual photons.

The Standard Model cross section ($f_4^V=f_5^V=0$) from \cite{jadach} shows 
agreement at the percent level with the one determined in Reference 
\cite{bardin}, where it is shown that all significant radiation effects 
come from ``universal'' radiator factors. This implies that an approach 
based on Equation \ref{eq:radiator} is justified in terms of the 
required precision.

\section*{The complete 
  $\bbox{\mathrm{\lowercase{e}^+ \lowercase{e}^-} \rightarrow 
  \mathrm{\lowercase{f}}\bar{\mathrm{\lowercase{f}}} 
  \mathrm{\lowercase{f}^\prime}\bar{\mathrm{\lowercase{f}^\prime}}}$ process}

  Additional non-resonant diagrams are taken into account in SM 
programs for general four-fermion production, like EXCALIBUR \cite{excalibur}. 
Under a reasonable set of kinematic cuts, the relative influence of those 
diagrams can be reduced, but not totally suppressed. This is due to the low 
cross section for resonant
ZZ production. Typical examples are those involving charged currents (relevant
in $\mathrm{e^+ e^-} \rightarrow \nu_e\overline{\nu}_e 
\mathrm{f}\bar{\mathrm{f}}$, 
$\mathrm{e^+ e^-} \rightarrow u\overline{d} d\overline{u}$, \ldots ) 
or multiperipheral effects in 
$\mathrm{e^+ e^-} \rightarrow \mathrm{e^+ e^-} \mathrm{f}\bar{\mathrm{f}}$. 
In addition, the influence of Fermi correlations in final states with identical fermions ($\mathrm{e^+ e^-} \rightarrow \mathrm{f}\bar{\mathrm{f}} 
\mathrm{f}\bar{\mathrm{f}}$) is unclear.

   In order to include these effects, the EXCALIBUR program has been 
extended.  All matrix elements from conversion diagrams with
two virtual Z particles $M_{\mathrm{Z}\mathrm{Z}}^{\rm EXC}
(\sigma,\lambda,\lambda^\prime;\bar{\Omega})$ are modified in the 
following way:
\begin{eqnarray}
\Delta^{\mathrm{Z}\mathrm{Z}}(\sigma,\lambda,\lambda^\prime;\bar{\Omega}) & 
\equiv &
\frac{\displaystyle \sum_{\lambda_{\mathrm {Z_1}},
   \lambda_{\mathrm {Z_2}}} M_{AC}~
                 M_{\mathrm {Z_1}\mathrm{f}\bar{\mathrm{f}}}~
M_{\mathrm {Z_2}\mathrm{f^\prime}\bar{\mathrm{f^\prime}}}}
        {\displaystyle \sum_{\lambda_{\mathrm {Z_1}},
\lambda_{\mathrm {Z_2}}} M_{\mathrm{Z}\mathrm{Z}}~
                 M_{\mathrm {Z_1}\mathrm{f}\bar{\mathrm{f}}}~
M_{\mathrm {Z_2}\mathrm{f^\prime}\bar{\mathrm{f^\prime}}}} \\
M_{\mathrm{Z}\mathrm{Z}}^{\rm EXC}(\sigma,\lambda,\lambda^\prime;\bar{\Omega}) 
& \rightarrow & 
M_{\mathrm{Z}\mathrm{Z}}^{\rm EXC}(\sigma,\lambda,\lambda^\prime;\bar{\Omega})~
\left(
1 + \Delta^{\mathrm{Z}\mathrm{Z}}(\sigma,\lambda,\lambda^\prime;\bar{\Omega}) 
\right) 
\end{eqnarray}

\noindent
where $M_{\mathrm{Z}\mathrm{Z}},M_{AC}, 
M_{\mathrm {Z_1}\mathrm{f}\bar{\mathrm{f}}}$ and 
$M_{\mathrm {Z_2}\mathrm{f^\prime}\bar{\mathrm{f^\prime}}}$ are the 
same terms defined in the NC08 approach. Based on this modification it is 
straightforward to define an alternative reweighting.
It will be identified as the ``FULL approach'' 
in the following. More detailed studies are reported in the next section.

\section*{The NC08 approach versus a full treatment}

\begin{figure}[htbp]
\begin{center}
\includegraphics[width=16.0truecm]{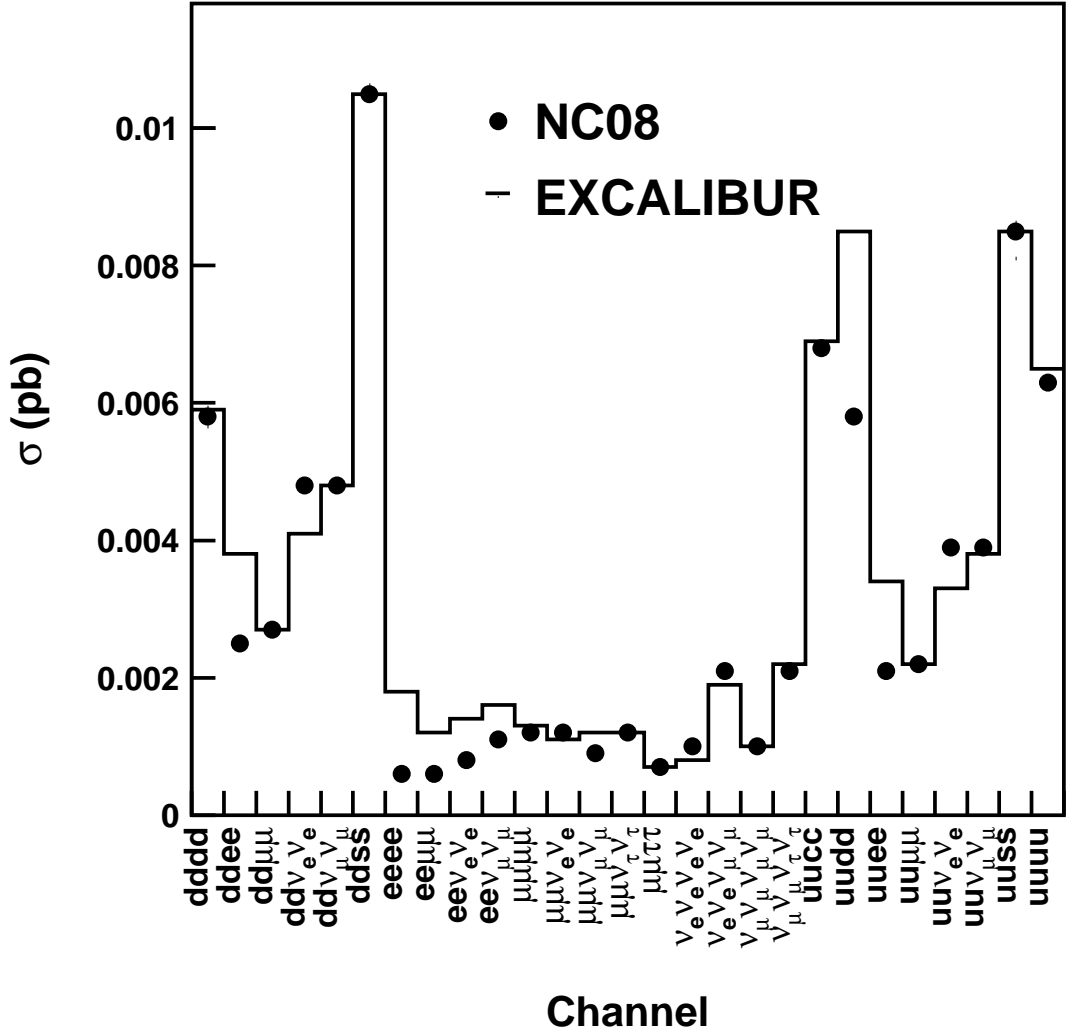}
\end{center}
\caption{Cross sections at $\sqrt{s}=190 \mathrm{\ Ge\kern -0.1em V}$ 
         for the different four-fermion 
         channels taking into account all the diagrams (histogram) and only
         the neutral conversion ones (NC08). Some cuts (described in the 
         text) have been applied in order to enhance the ZZ resonant
         contribution. Other diagrams are important when
         electrons, electronic neutrinos or fermions from
         the same isospin doublet are present in the final state.}
\label{fig:nc08_vs_exc}
\end{figure}

All checks presented in this section require a precise definition of
a ``$\mathrm{Z}\mathrm{Z}$ signal''. 
Channels involving electron or electronic neutrino
pairs in the final state have a non-negligible contribution
from non-conversion diagrams. Also final states with fermions 
from the same isospin doublet ($(\ell,\nu_\ell)$,(u,d),(c,s)) 
show a non-negligible charged current contribution. Therefore 
stringent cuts must be applied in order to select a sensible
$\mathrm{e^+ e^-} \rightarrow \mathrm{Z}\mathrm{Z}$ experimental signal. 
Our signal definition implies the following cuts: 
\begin{itemize}
\item The invariant masses of the two fermion-antifermion candidate pairs 
      must be in the range 70 GeV-105 GeV. 
\item In the final states with electrons, these electrons must verify 
      $|\cos \theta_{e}|<0.95$.
\item In the final states with WW contributions, the invariant masses of 
      the fermion pairs susceptible to come from W decay must be outside 
      the range 75 GeV-85 GeV. 
\end{itemize}
\noindent
  The SM cross sections within signal definition cuts at 
$\sqrt{s} = 190 \mathrm{\ Ge\kern -0.1em V}$ 
are shown in Figure~\ref{fig:nc08_vs_exc} for the different four-fermion channels. 
The EXCALIBUR generator is used. Two determinations are shown: one taking into 
account all the Standard Model diagrams and the other considering the 
neutral conversion diagrams only. Note that, 
in some cases, there are large differences between the two calculations. This 
already points to the convenience of using a full four-fermion approach, even
in the presence of strong cuts.

\begin{figure}[htb]
\begin{center}
\includegraphics[width=17.0truecm]{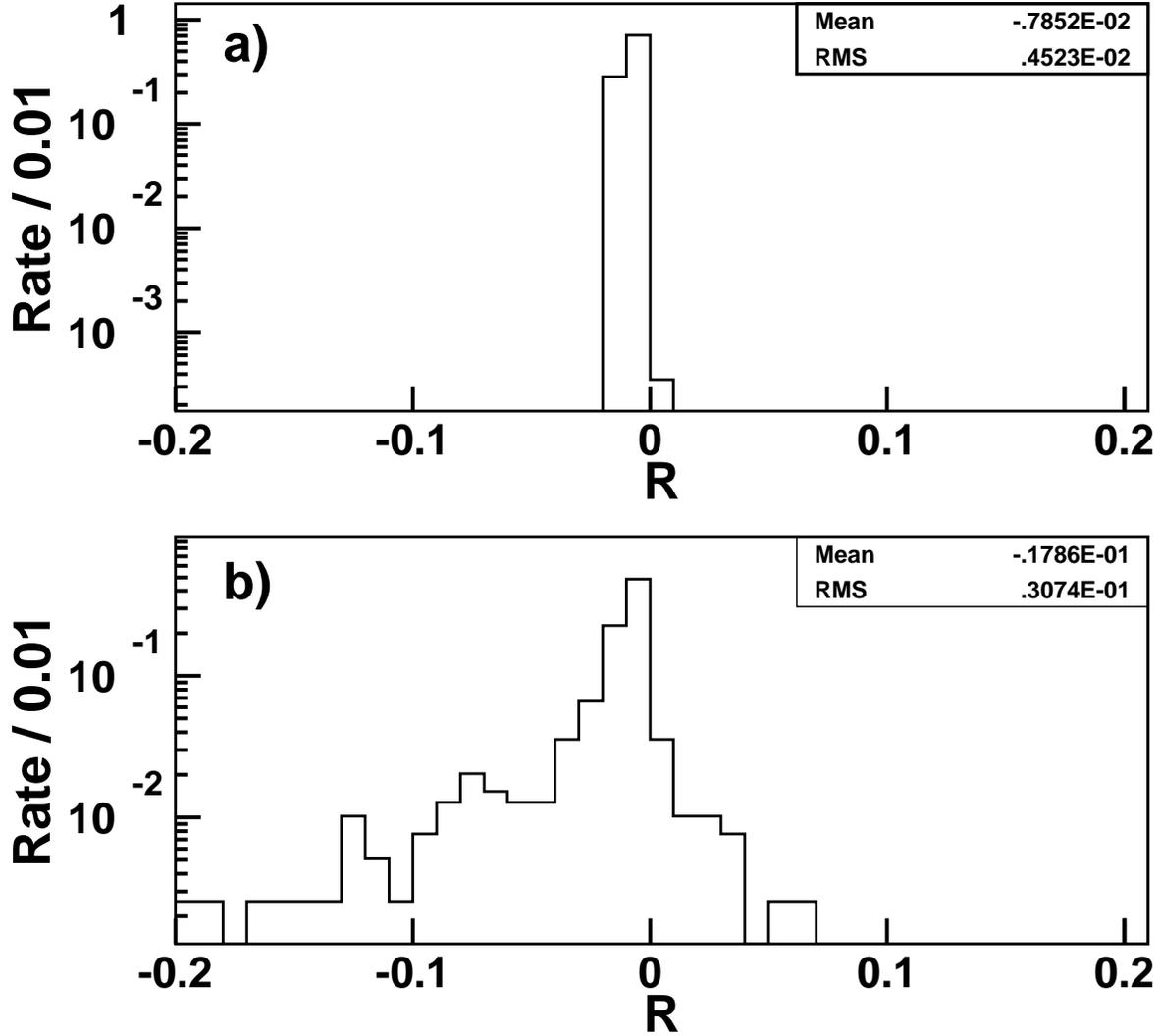}
\end{center}
\caption{Distributions of the relative difference, R, 
         between the SM squared matrix elements computed by EXCALIBUR and
         by the NC08 approach. 
         Only neutral conversion diagrams are taken into account. The plot 
         a) shows the case of final states without identical fermions 
       ($\mathrm{f}\bar{\mathrm{f}}\mathrm{f^\prime}\bar{\mathrm{f^\prime}}$).
         is shown in the plot.  
         The plot b) contains those final states with identical 
         fermions in the final state 
         ($\mathrm{f}\bar{\mathrm{f}} \mathrm{f}\bar{\mathrm{f}}$).
         The effect of Fermi correlations (no effort has been done to include 
         them in the NC08 approach) is visible as a tail in the 
         distribution.}
\label{fig:matrixel}
\end{figure}

  The next study is devoted to the SM matrix elements. For the 
same set of neutral conversion diagrams, the results obtained by 
EXCALIBUR and by the NC08 approach are compared. 
The relative differences are shown in Figure~\ref{fig:matrixel}.
Two groups are considered. The first group corresponds to all processes
without Fermi correlations, that is, those in which there are no identical 
fermions in the final state. A perfect agreement is observed in this case. 
The second group contains those processes in which there are identical 
particles in the final state. Let us note that no effort has been done in 
the NC08 approach to antisymmetrize the matrix element in the presence
of identical fermions. Although the effect is understood and it can be 
trivially included, we want to show that it is not totally negligible. It 
may exceed $10\%$ for some phase space configurations.

\begin{figure}[htb]
\begin{center}
\includegraphics[width=18.0truecm]{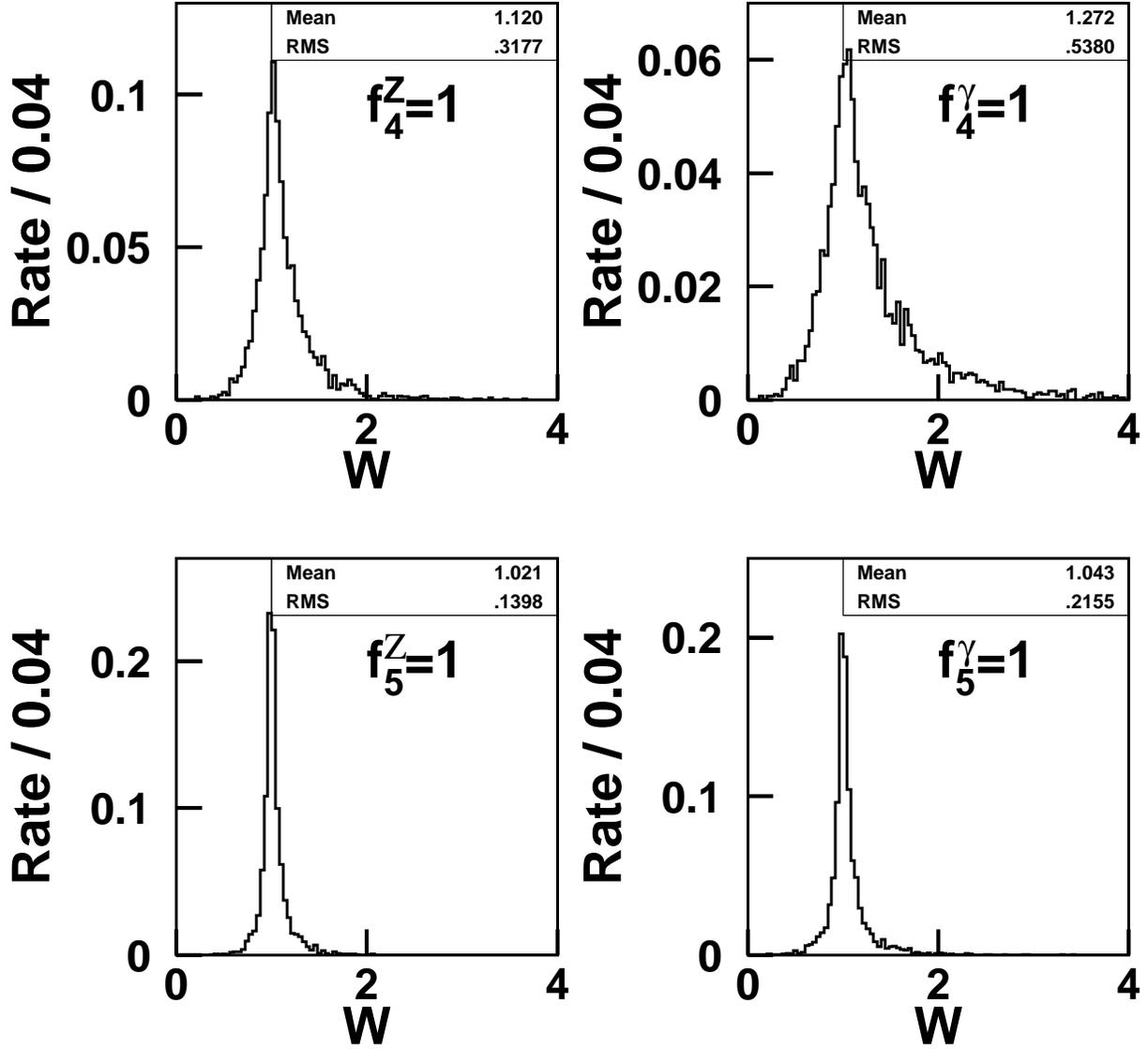}
\end{center}
\caption{Distributions of the reweighting factor, W, in the 
         $\mathrm{e^+ e^-} \rightarrow \mathrm{Z}\mathrm{Z} (\gamma) 
         \rightarrow \mathrm{f}\bar{\mathrm{f}} 
         \mathrm{f^\prime}\bar{\mathrm{f^\prime}} (\gamma)$
         process for different values of anomalous ZZV couplings. The 
         reweighting factors are obtained with the FULL
         approach at $\sqrt{s}=190 \mathrm{\ Ge\kern -0.1em V}$. 
         Only neutral conversion diagrams are taking into account.}
\label{fig:check2}
\end{figure}

\begin{figure}[htb]
\begin{center}
\includegraphics[width=18.0truecm]{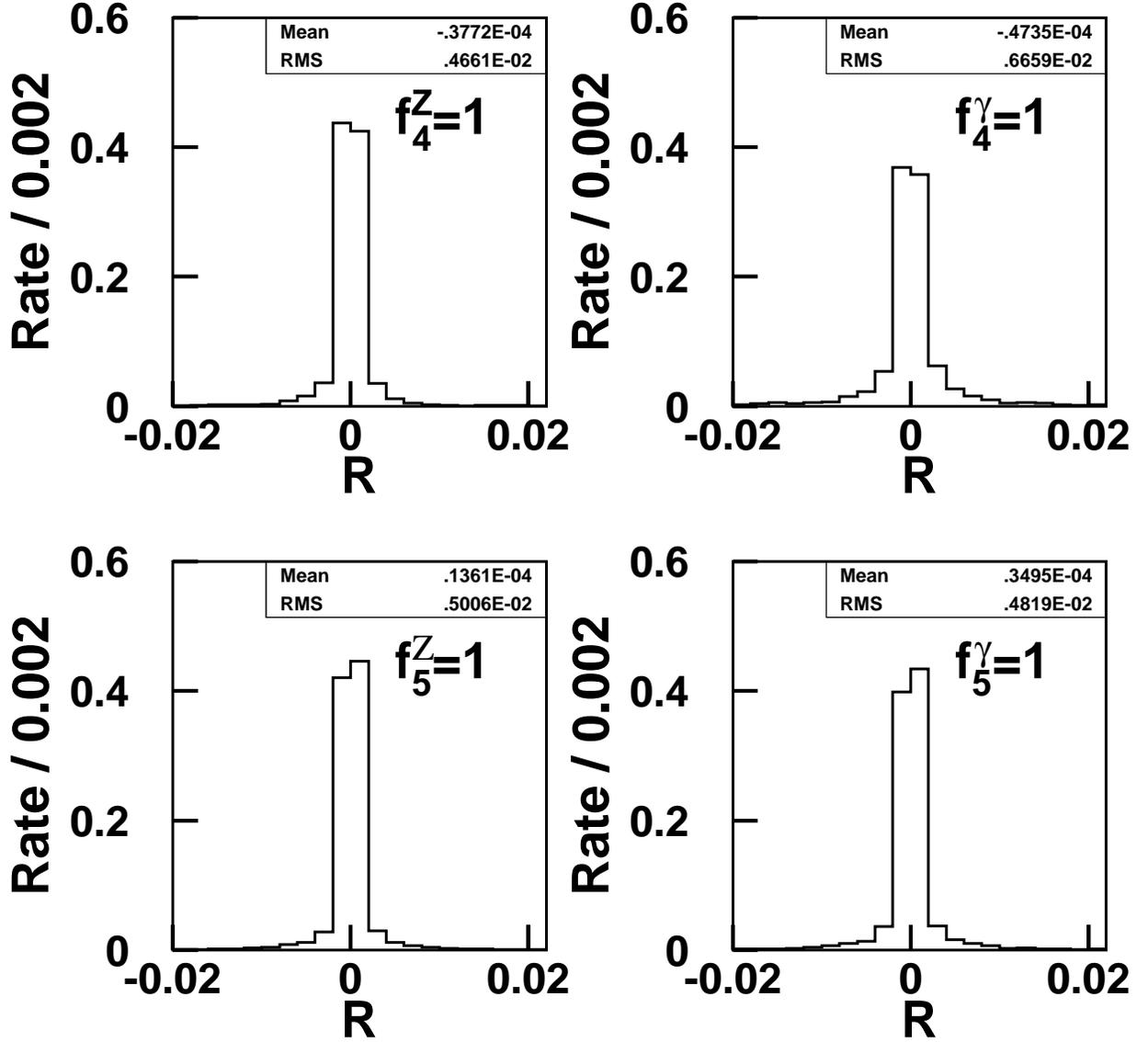}
\end{center}
\caption{Distributions of the relative difference, R, between the weights 
         computed with the FULL and the NC08 approaches. The
         process is $\mathrm{e^+ e^-} \rightarrow 
         \mathrm{Z}\mathrm{Z} (\gamma) \rightarrow 
         \mathrm{f}\bar{\mathrm{f}} 
         \mathrm{f^\prime}\bar{\mathrm{f^\prime}} (\gamma)$ at
         $\sqrt{s}=190 \mathrm{\ Ge\kern -0.1em V}$ for different values of 
         anomalous ZZV couplings. 
         Only neutral conversion diagrams are taking into account.}
\label{fig:check3}
\end{figure}

   Figure \ref{fig:check2} shows the distribution of weights at 
$\sqrt{s}=190 \mathrm{\ Ge\kern -0.1em V}$ for non-zero values of the 
anomalous couplings.
The fact that the distributions are not extremely narrow 
indicates the presence of effects other than just an excess of events. 
To compare the implementations in the presence of anomalous couplings, the 
Standard Model distributions are reweighted according to the NC08 and 
FULL approaches. Again, only neutral conversion diagrams 
are considered. The relative differences between the weights
assigned by the two approaches
are shown in Figure~\ref{fig:check3}. There is agreement at the percent level.
Fermi correlations in the final state have a small effect on the 
weights. The reason could be related to the fact that the discrepancy 
factorizes in a similar way SM and anomalous terms.

The last study evaluates the influence of non-conversion diagrams
in the presence of anomalous couplings. The averages of the reweighting factors 
for the FULL and the NC08 approaches 
are compared in tables~\ref{tab:fz}- \ref{tab:fg}. The set of 
cuts defining the ZZ resonant region is applied in all cases.  The differences 
are typically below 10\%, but not negligible. This points again to the 
convenience of considering all possible diagrams contributing to the 
$\mathrm{e^+ e^-} \rightarrow \mathrm{f}\bar{\mathrm{f}} 
\mathrm{f^\prime}\bar{\mathrm{f^\prime}}$ process.

\section*{Measurement of $\bbox{\mathrm{Z}\mathrm{Z}\mathrm{Z}}$ and 
            $\bbox{\mathrm{Z}\mathrm{Z}\gamma}$ anomalous couplings}

  To determine the values of the anomalous couplings from the data, 
the histogram of the most relevant variable for each four-fermion channel may
be used. The following binned likelihood function is then maximized:
\begin{eqnarray}
\log({\cal L}) = \sum^{N_{bin}}_{j=1} 
\left[ N_{data}(j) \log N_{expected}(j;f_i^V) - N_{expected}(j;f_i^V) \right]
\end{eqnarray}

  The expected number of events is computed as:
$N_{expected}(j;f_i^V) = N_{signal}(j;f_i^V) + N_{background}(j)$. The
background contribution does not depend on the anomalous couplings. The
signal contribution includes all four-fermion final states compatible 
with the exchange of two $\mathrm{Z}$ bosons. It is computed by reweighting the 
Standard Model distributions with the FULL approach and 
taking into account the $f_i^V$ values of the anomalous couplings.

  This method has been 
applied in the analysis of 
$\mathrm{e^+ e^-} \rightarrow \mathrm{f}\bar{\mathrm{f}} 
\mathrm{f^\prime}\bar{\mathrm{f^\prime}}$ final states by 
the L3 Collaboration~\cite{zzpapers}. The discriminating variables are
the invariant masses of the lepton pairs in leptonic decays and neural net 
outputs for hadronic decays. They obtain the following 95\% confidence level
limits on the existence of anomalous $\mathrm{Z}\mathrm{Z}$V couplings:
\begin{displaymath}
-1.9  \leq f_4^{\mathrm{Z}}    \leq 1.9;\   \ 
-5.0  \leq f_5^{\mathrm{Z}}    \leq 4.5;\   \
-1.1  \leq f_4^{\gamma} \leq 1.2;\   \ 
-3.0  \leq f_5^{\gamma} \leq 2.9.
\end{displaymath}

   As an example of the use of more sensitive variables at higher energies 
we will consider the semileptonic process $\mathrm{e^+ e^-} 
\rightarrow \mathrm{\ell^+ \ell^-} \mathrm{q}\bar{\mathrm{q}}$ at 
$\sqrt{s}=500 \mathrm{\ Ge\kern -0.1em V}$. This channel is expected to 
provide a clean signature for $\mathrm{Z}\mathrm{Z}$ production
if the Higgs mass is away from the Z mass region.
In order to enhance the ZZ signal component, the invariant mass of the leptons 
is required to be in the 70 GeV to 150 GeV range, the recoiling hadronic 
mass has to be larger than 50 GeV and the polar angle of electrons
and positrons, $\theta_{\mathrm e}$, must satisfy
$\mid \cos\theta_{\mathrm e}\mid < 0.8$. The differential cross section of the 
process in the presence of an anomalous coupling $f$ can be expressed as 
follows:

\begin{eqnarray}
   \left. \frac{d^2 \sigma}{d(O_1)~d(O_2)} \right|_f= 
   \left. \frac{d^2 \sigma}{d(O_1)~d(O_2)} \right|_{f=0}~
              \left( 1+ f~O_1 + f^2~O_2 \right) 
\end{eqnarray}

\begin{figure}[htb]
\begin{center}
\includegraphics[width=18.0truecm]{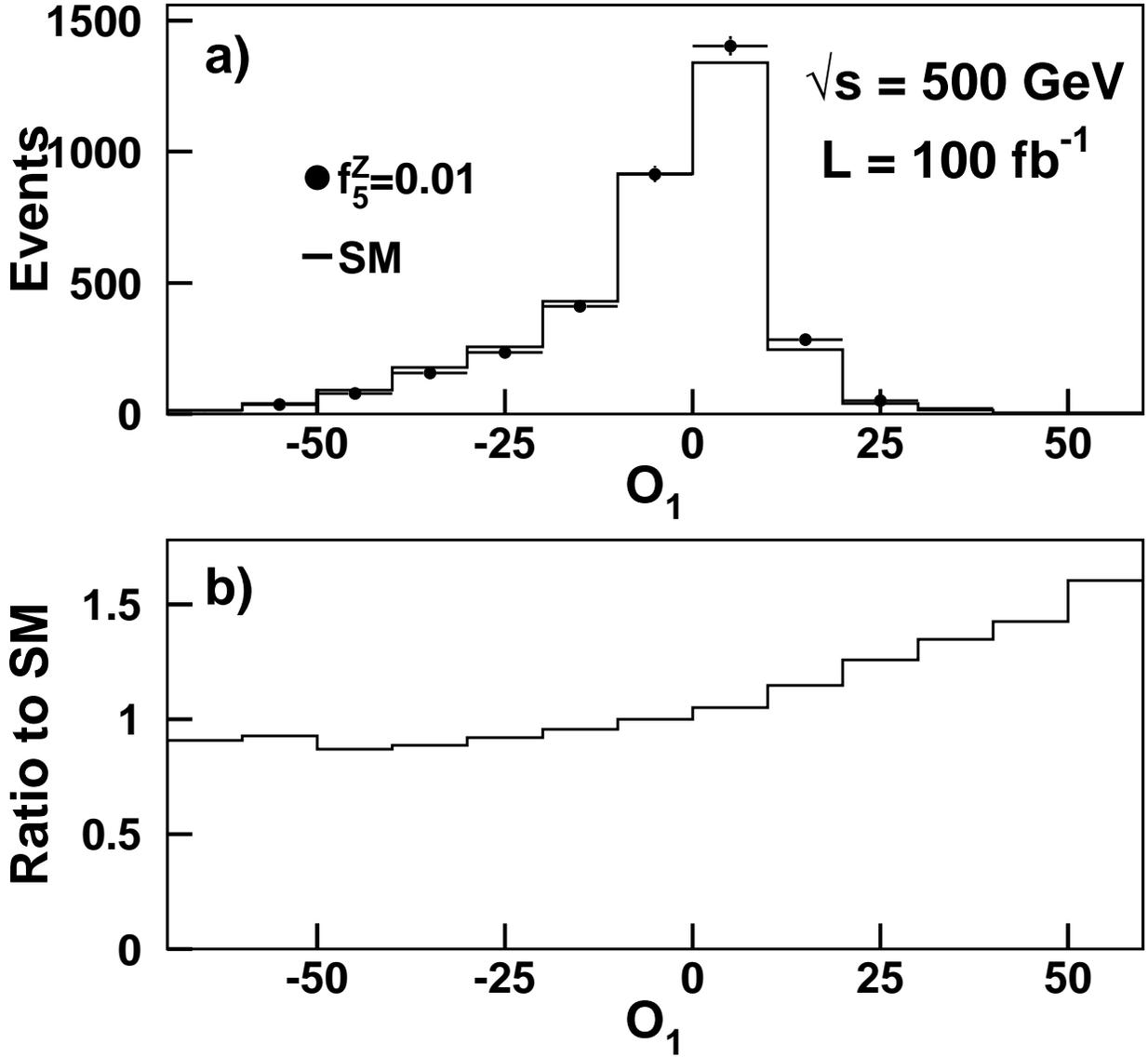}
\end{center}
\caption{Distributions of the optimal variable $O_1$ in the Standard Model
         (histogram) and in the presence of an anomalous coupling 
         $f_5^{\mathrm{Z}}=0.01$ (points). The process 
         under consideration is $\mathrm{e^+ e^-} \rightarrow 
         \mathrm{\ell^+ \ell^-} \mathrm{q}\bar{\mathrm{q}}$ at
         $\sqrt{s}=500 \mathrm{\ Ge\kern -0.1em V}$ and for an 
         integrated luminosity of 100 \mbox{fb$^{-1}$}. The second plot shows 
         the ratio between the anomalous and the SM distributions.
         A fit to the anomalous distribution leads to the value 
         $f_5^{\mathrm{Z}} = 0.010 \pm 0.002$.}
\label{fig:oo1}
\end{figure}

   The variables $O_1$ and $O_2$ are functions of the phase space variables 
of an event. They are independent of $f$. 
 The previous equation guarantees that the maximal information on 
$f$ is obtained by a study of the event density as a function of the variables 
$O_1$ and $O_2$. These variables are usually called ``optimal observables''. 
Given an event characterized by the phase space point $\bar{\Omega}$, a simple 
expression for the optimal observables is:

\begin{eqnarray}
   O_1(\bar{\Omega}) & = & \frac{W(\bar{\Omega};f=+1) - 
                               W(\bar{\Omega};f=-1)}{2} \\
   O_2(\bar{\Omega}) & = & \frac{W(\bar{\Omega};f=+1) + 
                               W(\bar{\Omega};f=-1)}{2} - 1
\end{eqnarray}

\noindent
where $W(\bar{\Omega};f)$ is the reweighting factor used to transform SM 
distributions into anomalous distributions at $\bar{\Omega}$ for an anomalous 
coupling $f$. At $\sqrt{s}=500 \mathrm{\ Ge\kern -0.1em V}$ and for small 
values of $f$ the dominant term in the ZZ differential cross 
section is the one associated to $O_1$. In real life, the exact values of the 
phase space variables are unknown, but we may use as an approximation the 
phase space variables reconstructed by the detector. For this exercise we
assume energy resolutions of $5\%$ for quarks and leptons and a jet angular 
resolution of 40 mrad. The weight is symmetrized under the interchange 
of quark types, assuming that the quark flavour can not be identified. The SM 
distribution of the variable $O_1$ for an integrated luminosity of 100 
\mbox{fb$^{-1}$} is shown in 
Figure \ref{fig:oo1}, together with the effect of an anomalous coupling 
$f_5^{\mathrm{Z}}=0.01$. Note that the ratio between the anomalous and the SM 
distributions is not constant. This is a demonstration of the 
sensitivity of the variable and of the existence of anomalous effects different
from a simple change in the total cross section. From a likelihood fit to the 
anomalous distribution we obtain a value of $f_5^{\mathrm{Z}}=0.010 \pm 0.002$. 
Note the increase in sensitivity as compared to LEP2 present limits.

\section*{Acknowledgements}
   We would like to thank J. Biebel and A. Felbrich for useful discussions 
and cross-checks. We are also grateful to R. Pittau for providing the last 
version of the EXCALIBUR program for L3. We specially thank the help, 
positive criticism and support from our L3-ZZ collaborators during this time.

\appendix
\section{Most General Anomalous ZZV Couplings} \label{app:a}

     The most general Lorentz invariant anomalous coupling structure for the 
$\mathrm{Z}\mathrm{Z} V$ vertex function is, similarly to 
the WWV case \cite{hagiwara_et_al}:
\begin{eqnarray}
\Gamma^{\alpha\beta\mu}_{\mathrm {Z_1}\mathrm {Z_2} V} = & 
    \frac{s-m_V^2}{ m_{\mathrm{Z}}^2} \left\{ \right. 
  i f_4^V~( (q_{\mathrm {Z_1}}+q_{\mathrm {Z_2}})^\alpha~g^{\mu\beta} +
               (q_{\mathrm {Z_1}}+q_{\mathrm {Z_2}})^\beta~g^{\mu\alpha} ) 
  +~i f_5^V~\epsilon^{\alpha\beta\mu\rho}~
   (q_{\mathrm {Z_1}}-q_{\mathrm {Z_2}})_\rho 
            \left. \right\} \nonumber \\
    & + \nonumber \\
                                   & \frac{s-m_V^2}{ m_{\mathrm{Z}}^2}
   ~\frac{(q_{\mathrm {Z_1}}^2-q_{\mathrm {Z_2}}^2)}{\Lambda^2} \left\{ \right. 
     f_1^V~(q_{\mathrm {Z_1}}-q_{\mathrm {Z_2}})^\mu~g^{\alpha\beta} 
   -~\frac{f_2^V}{ m_{\mathrm{Z}}^2}~(q_{\mathrm {Z_1}}-q_{\mathrm {Z_2}})^\mu~
              (q_{\mathrm {Z_1}}+q_{\mathrm {Z_2}})^\alpha
           ~(q_{\mathrm {Z_1}}+q_{\mathrm {Z_2}})^\beta \nonumber \\
 & +~f_3^V~( (q_{\mathrm {Z_1}}+q_{\mathrm {Z_2}})^\alpha~g^{\mu\beta} -
               (q_{\mathrm {Z_1}}+q_{\mathrm {Z_2}})^\beta~g^{\mu\alpha} ) 
  -~f_6^V~\epsilon^{\alpha\beta\mu\rho}~
          (q_{\mathrm {Z_1}}+q_{\mathrm {Z_2}})_\rho \nonumber \\
  & -~\frac{f_7^V}{ m_{\mathrm{Z}}^2}~
         (q_{\mathrm {Z_1}}-q_{\mathrm {Z_2}})^\mu~
         \epsilon^{\alpha\beta\rho\sigma}~
         (q_{\mathrm {Z_1}}+q_{\mathrm {Z_2}})_\rho~
          (q_{\mathrm {Z_1}}-q_{\mathrm {Z_2}})_\sigma \left. \right\}
\end{eqnarray}

    The global factor $(s-m_V^2)/ m_{\mathrm{Z}}^2$ is introduced by 
convention~\footnote{The presence of $ m_{\mathrm{Z}}$ in the denominator is 
arbitrary. It allows the introduction of a dimensionless coupling constant 
without adding new unknown scale parameters. A physically more motivated 
choice is to substitute $ m_{\mathrm{Z}}$ by a scale of new physics $\Lambda$. 
In this way, the unknown coupling is of the type $f/\Lambda^2$, with $f$ of 
order unity, but the higher dimensionality of the term is exhibited.} 
in order to preserve gauge invariance when 
$V=\gamma$ and Bose-Einstein symmetry for $V=\mathrm{Z}$ in the on-shell limit.
 Note that only the terms associated to $f_4^V$ and $f_5^V$ survive in 
the limit in which both $\mathrm{Z}$ are on-shell.

The $\mathrm{Z}\mathrm{Z} V$ vertex function must be symmetric under the 
interchange $(\mathrm {Z_1},\alpha) \leftrightarrow (\mathrm {Z_2},\beta)$. 
For the terms associated 
to $f_1^V, f_2^V, f_3^V, f_6^V$ and $f_7^V$ this requirement forces 
the presence of an additional Lorentz invariant factor:
any antisymmetric function of $(q_{\mathrm {Z_1}}^2-q_{\mathrm {Z_2}}^2)$. 
Our minimal choice is: $(q_{\mathrm {Z_1}}^2-q_{\mathrm {Z_2}}^2)/\Lambda^2$, 
where $\Lambda$ is a scale related to new physics. 
Two comments are in order here:

\begin{itemize}
\item The anomalous couplings $f_1^V, f_2^V, f_3^V, f_6^V, f_7^V$ are 
      necessarily associated to Lagrangians of higher dimension than those 
      associated to $f_4^V, f_5^V$.
       
\item The sensitivity to the anomalous $f_1^V, f_2^V, f_3^V, f_6^V, f_7^V$
      couplings is further reduced due to the relatively small size of the 
      Z width: $(q_{\mathrm {Z_1}}^2-q_{\mathrm {Z_2}}^2)/\Lambda^2 \approx 
      {\cal O}(\Gamma_{\mathrm{Z}} m_{\mathrm{Z}}/\Lambda^2)$.
\end{itemize}

     The existence of additional $(q_i^2- m_{\mathrm{Z}}^2)$ factors when 
at least one of the final $\mathrm{Z}$ is off-shell has also been noticed in 
Reference \cite{hagiwara_et_al}. For completeness, we provide the full 
expressions for all possible anomalous couplings matrix elements, 
$M_{AC}^{f_i^V} \equiv M_{AC}^{f_i^V}
 (\sigma,\overline{\sigma},\lambda_{\mathrm {Z_1}},\lambda_{\mathrm {Z_2}})$:
\begin{eqnarray}
   M_{AC}^{f_4^V} = & -i~e f_4^V g_\sigma^{Vee}~\frac{s}{ m_{\mathrm{Z}}^2}~
   \delta_{\sigma,-\overline{\sigma}} \left [
    \epsilon_{\mathrm {Z_1}}^{0*}
    (\epsilon_{\mathrm {Z_2}}^{1*} + i\sigma \epsilon_{\mathrm {Z_2}}^{2*}) +
    \epsilon_{\mathrm {Z_2}}^{0*}
    (\epsilon_{\mathrm {Z_1}}^{1*} + i\sigma \epsilon_{\mathrm {Z_1}}^{2*}) 
           \right ] \\
   M_{AC}^{f_5^V} = & -i~e f_5^V g_\sigma^{Vee}~
         \frac{\sqrt{s}}{ m_{\mathrm{Z}}^2}~
   \delta_{\sigma,-\overline{\sigma}}~
    (\epsilon^{1\alpha\beta\rho} + i\sigma \epsilon^{2\alpha\beta\rho} )~
    \epsilon_{\mathrm {Z_1}\alpha}^{*}~\epsilon_{\mathrm {Z_2}\beta}^{*}~
         (q_{\mathrm {Z_1}\rho}-q_{\mathrm {Z_2}\rho}) \\
   M_{AC}^{f_1^V} = & -e f_1^V g_\sigma^{Vee}~\frac{\sqrt{s}}{ m_{\mathrm{Z}}^2}
    ~\frac{(q_{\mathrm {Z_1}}^2 - q_{\mathrm {Z_2}}^2)}{\Lambda^2}~
   \delta_{\sigma,-\overline{\sigma}} 
    (\epsilon_{\mathrm {Z_1}}^{*} \epsilon_{\mathrm {Z_2}}^{*})~
    \left[ (q_{\mathrm {Z_1}}^{1}-q_{\mathrm {Z_2}}^{1}) + i\sigma 
         (q_{\mathrm {Z_1}}^{2}-q_{\mathrm {Z_2}}^{2})
    \right]            \\
   M_{AC}^{f_2^V} = &  e f_2^V g_\sigma^{Vee}~\frac{s^{3/2}}{ m_{\mathrm{Z}}^4}
    ~\frac{(q_{\mathrm {Z_1}}^2 - q_{\mathrm {Z_2}}^2)}{\Lambda^2}~
   \delta_{\sigma,-\overline{\sigma}} 
    \epsilon_{\mathrm {Z_1}}^{0*}~\epsilon_{\mathrm {Z_2}}^{0*}~
    \left[ (q_{\mathrm {Z_1}}^{1}-q_{\mathrm {Z_2}}^{1}) + i\sigma 
         (q_{\mathrm {Z_1}}^{2}-q_{\mathrm {Z_2}}^{2})
    \right]            \\
   M_{AC}^{f_3^V} = & -e f_3^V g_\sigma^{Vee}~\frac{s}{ m_{\mathrm{Z}}^2}
    ~\frac{(q_{\mathrm {Z_1}}^2 - q_{\mathrm {Z_2}}^2)}{\Lambda^2}~
   \delta_{\sigma,-\overline{\sigma}}~\left[
    \left( \epsilon_{\mathrm {Z_1}}^{0} (\epsilon_{\mathrm {Z_2}}^{1}+i\sigma 
          \epsilon_{\mathrm {Z_2}}^{2}) \right) - 
    \left( \epsilon_{\mathrm {Z_2}}^{0} (\epsilon_{\mathrm {Z_1}}^{1}+i\sigma 
          \epsilon_{\mathrm {Z_1}}^{2}) \right) \right] \\
   M_{AC}^{f_6^V} = &  e f_6^V g_\sigma^{Vee}~\frac{s}{ m_{\mathrm{Z}}^2}
    ~\frac{(q_{\mathrm {Z_1}}^2 - q_{\mathrm {Z_2}}^2)}{\Lambda^2}~
   \delta_{\sigma,-\overline{\sigma}}~
    (\epsilon^{1\alpha\beta 0} + i\sigma \epsilon^{2\alpha\beta 0} )~
    \epsilon_{\mathrm {Z_1}\alpha}^{*} \epsilon_{\mathrm {Z_2}\beta}^{*} \\
   M_{AC}^{f_7^V} = &  e f_7^V g_\sigma^{Vee}
       ~\frac{s}{ m_{\mathrm{Z}}^4}~
      \frac{(q_{\mathrm {Z_1}}^2 - q_{\mathrm {Z_2}}^2)}{\Lambda^2}~
   \delta_{\sigma,-\overline{\sigma}}~
    \left[ (q_{\mathrm {Z_1}}^{1}-q_{\mathrm {Z_2}}^{1}) + i\sigma 
        (q_{\mathrm {Z_1}}^{2}-q_{\mathrm {Z_2}}^{2}) \right]
    ~\epsilon^{\alpha\beta 0\rho}~(q_{\mathrm {Z_1}}-q_{\mathrm {Z_2}})_\rho
    ~\epsilon_{\mathrm {Z_1}\alpha}^{*} \epsilon_{\mathrm {Z_2}\beta}^{*}
\end{eqnarray}


\begin{table} [htp]
\begin{center}
\begin{tabular}{c|cc|cc}
 & \multicolumn{2}{c|}{$f_4^{\mathrm{Z}} = 1$} & 
   \multicolumn{2}{c}{$f_5^{\mathrm{Z}} = 1$} \\
\hline
Final state & NC08 & FULL & NC08 & FULL \\
\hline
$\rm q   \bar q   q   \bar q   $  & 1.128     &  1.123
                                  & 1.023     &  1.022    \\
$\rm q   \bar q   \nu \bar \nu $  & 1.131     &  1.150
                                  & 1.019     &  1.024    \\
$\rm q   \bar q   l   \bar l   $  & 1.105     &  1.091
                                  & 1.013     &  1.006    \\
$\rm l   \bar l   \nu \bar \nu $  & 1.101     &  1.082
                                  & 1.025     &  1.013    \\
$\rm l   \bar l   l   \bar l   $  & 1.128     &  1.054
                                  & 1.024     &  1.007    \\
$\rm \nu \bar \nu \nu \bar \nu $  & 1.084     &  1.170
                                  & 1.022     &  1.036    \\
\end{tabular}       
\caption{
Average reweighting factors in the presence of anomalous ZZZ couplings. 
The case with only neutral conversion diagrams (NC08) is compared to 
the complete approach with all Feynman diagrams (FULL).} 
\label{tab:fz}
\end{center}
\end{table}

\begin{table} [htp]
\begin{center}
\begin{tabular}{c|cc|cc}
 & \multicolumn{2}{c|}{$f_4^\gamma = 1$} & 
                         \multicolumn{2}{c}{$f_5^\gamma = 1$} \\
\hline
Final state &  NC08 &  FULL & NC08 & FULL  \\
\hline
$\rm q   \bar q   q   \bar q   $  & 1.350     &  1.336
                                  & 1.041     &  1.041    \\
$\rm q   \bar q   \nu \bar \nu $  & 1.339     &  1.385
                                  & 1.047     &  1.050    \\
$\rm q   \bar q   l   \bar l   $  & 1.273     &  1.235
                                  & 1.032     &  1.026    \\
$\rm l   \bar l   \nu \bar \nu $  & 1.218     &  1.181
                                  & 1.031     &  1.004    \\
$\rm l   \bar l   l   \bar l   $  & 1.380     &  1.161
                                  & 1.033     &  1.019    \\
$\rm \nu \bar \nu \nu \bar \nu $  & 1.254     &  1.463
                                  & 1.071     &  1.082    \\
\end{tabular}       
\caption{
Average reweighting factors in the presence of anomalous 
  $\mathrm{Z}\mathrm{Z}\gamma$ couplings.  
 The case with only neutral conversion diagrams (NC08) is compared 
to the complete approach with all Feynman diagrams (FULL).} 
\label{tab:fg}
\end{center}
\end{table}

\newpage

\newpage

\newpage

\newpage

\newpage

\newpage

\newpage

\newpage

\newpage

\newpage

\newpage

\end{document}